\documentstyle[preprint,aps,pre,eqsecnum]{revtex}
\input epsf
\tighten
\draft
\begin{document}
\title{Dynamical Relaxation and Universal Short-Time Behavior\\
in Finite Systems: The Renormalization Group Approach}\vspace{25mm}
\author{U.\ Ritschel and H.\ W.\ Diehl}
\address{Fachbereich Physik, Universit\"at - Gesamthochschule -  Essen,\\
D-45117 Essen,
Federal Republic of Germany}
\date{submitted to Nucl. Phys. B}
\maketitle
\begin{abstract}
We study how the finite-sized $n$-component model A with
periodic boundary conditions relaxes near its bulk critical point
from an initial nonequilibrium state with short-range correlations.
Particular attention is paid to the universal long-time traces
that the initial condition leaves.
An approach based on renormalization-group
improved perturbation theory in $4-\epsilon$ space dimensions
and a nonperturbative treatment of the $\bbox{q}=\bbox{0}$ mode
of the fluctuating order-parameter field is developed.
This leads to a
renormalized effective stochastic equation
for this mode in the background of the other,
$\bbox{q}\ne\bbox{0}$ modes;
we explicitly derive it to one-loop order,
show that it  takes the expected finite-size
scaling form at the fixed point,
and solve it numerically. Our results confirm
for general $n$ that the amplitude of
the magnetization density $m(t)$
in the linear relaxation-time regime
depends
on the
initial magnetization in the universal fashion originally found in our
large-$n$ analysis
[J.\ Stat.\ Phys.\ 73 (1993) 1].
The anomalous short-time power-law increase of $m(t)$ also is recovered.
For $n=1$, our results are in fair agreement
with  recent Monte Carlo simulations
by  Li, Ritschel, and Zheng
[J.\ Phys.\ A 27 (1994) L837] for the three-dimensional Ising model.
\end{abstract}

\pacs{PACS numbers: 64.60.Ht, 05.70.Jk, 05.70.Ln, 75.40.Gb}

\narrowtext
\section{Introduction}
In the classical work on critical dynamics
\cite{halp,hoha,domi,firacz,baja}
the important role of  initial conditions for relaxation processes
was not fully appreciated.
Guided by the (incorrect) idea that all
memory of the initial state would be lost after
a {\it microscopic} time span, so that
the initial condition would leave no universal
traces on {\it macroscopic} time scales, most authors
refrained from a thorough analysis of the effects of the
initial condition, taking the relaxation process
to start in the infinite past $t=-\infty$.
Others chose initial conditions at time $t=0$ and
incorporated the initial value $m_i=m(0)$ of
the order-parameter density $m(t)$ in the
scaling analysis of the relaxation process,
but tacitly presumed that $m_i$
scaled with the same dimension as $m(t>0)$ \cite{hoha,gold1,gold}.

As shown some time ago by Janssen et al.\ \cite{jans}, this
assumption is in general incorrect.
Due to its failure,
quantities relaxing from an
initial nonequilibrium state with short-range correlations
may exhibit interesting
short-time anomalies on {\it macroscopic} time scales.
To be specific, let us consider a $d$-dimensional
Ising system with purely relaxational dynamics,
which relaxes at the Curie temperature
$T=T_c$ from an initial nonequilibrium state
prepared by a rapid quench
from a thermal equilibrium state with $T\gg T_c$ and small $m_i$.
We assume that $d$ is below the upper
critical dimension $d^*$ ($=4$), but above the lower one, $d_*$ ($=2$).
The following is now well-established
\cite{jans,janszep,own,own2,mont,gras}: (i) The initial value
$m_i$ gives rise to a
{\it macroscopic} time scale $t_i\sim m_i^{-z/x_i}$,
where $z$ and $x_i$ denote the usual dynamic bulk exponent and
the scaling dimension of $m_i$, respectively.
(ii) The latter is greater
than $x_\phi\equiv \beta/\nu$, the scaling dimension of the
equilibrium order parameter $m(\infty)$, so that
\begin{equation}\label{theta}
\theta'=\frac{x_i-x_{\phi}}{z}
\end{equation}
is positive. (iii) For $t\lesssim t_i$, $m(t)$
displays an anomalous increase
of the form $\sim t^{\theta'}$, which crosses over to
the familiar nonlinear relaxational behavior
$\sim t^{-\beta/\nu z}$ for $t\gtrsim t_i$.

Similar short-time anomalies have been found
for other dynamic universality classes
\cite{oerd} and in the case of
relaxation near a tricritical point
\cite{oerd2}.

On a qualitative level, the puzzling fact that $m(t)$
actually {\it increases}
in the above short-time regime even though the equilibrium
state that is approached as $t\to\infty$ has zero magnetization
can be understood as follows  (cf.\ \cite{janszep}).
By choice of the initial state,
its correlation length $\xi(t\!=\!0)$ is small. As $t$ grows,
$\xi(t)$ will increase, too. Fluctuation effects
will be suppressed as long as $\xi(t)$ is small. Let us suppose
for a moment that fluctuation
effects can be neglected completely for small $t$. Then a
mean-field (MF) description should be valid.
For systems with short-range interactions this is known to
give a critical temperature $T_c^{\text{MF}}$ significantly
higher than the exact $T_c$. Hence for short times the system
should behave as if it were in the ordered MF phase,
with $m(t)$ trying to approach the corresponding
nonzero value of the spontaneous MF magnetization.
In reality, fluctuations must {\it not} be neglected when $d<d^*$,
not even in the short-time regime. This is clearly borne out by the
renormalization group (RG) analyses \cite{jans,oerd},
which show that loop corrections {\it do}
contribute to the short-time behavior,
producing nontrivial short-time exponents $\theta'$ and $x_i$.
Consequently, one must first integrate out all degrees of freedom
with wave lengths up to the scale $\xi(t)$ using the RG
before reliable information can be extracted by perturbative means.
This necessity of taking into account
fluctuations will lead to quantitative modifications.
Yet it seems reasonable to expect
that the tendency of $m(t)$ to
increase at short times should qualitatively persist.
Accepting this argument one concludes that
as $t$ and $\xi(t)$ grow,
the system will eventually realize that
the true $T_c$ is much lower and {\it not} below the
temperature $T$ of the surrounding heat bath.
Then $m(t)$ will start its decrease towards zero.

Recently we have
extended the work of Janssen et al.\ \cite{jans} to
systems of finite size \cite{own,own2}.
Specifically, a cube of large, but finite, side length
$L$ was considered under periodic boundary conditions.
As is well known, $L$ gives rise to a
characteristic time of the form $t_L\sim L^z$ \cite{sanch,ZJ,niel,fsli}.
At bulk criticality, this is the only other macroscopic
time scale besides $t_i$. In the case of the dynamic universality
class considered, namely the one represented by the familiar
model A of Refs.\ \onlinecite{halp,hoha},
the asymptotic long-time behavior is of the exponentially
decaying form
\begin{equation}\label{amde}
m(t\gg t_L)\approx M_{\infty}\,{\rm e}^{-t/t_L}\,,
\end{equation}
which identifies $t_L$ as the linear relaxation time at $T=T_c$.
While $t_L$ is independent of the initial state, the amplitude
$M_\infty$
was found to exhibit a {\it universal}
dependence on both $m_i$ and $L$. It may be written as \cite{own2}
\begin{equation}\label{minf}
M_{\infty}(L,\, m_i)\approx  L^{-\beta/\nu } \>
{\cal G}(t_i/t_L)\;,
\end{equation}
where ${\cal G}$ is a universal function, up to redefinitions
of its overall amplitude and the scale of its argument $w=t_i/t_L$.
Since ${\cal G}$ approaches a constant
in the limit $w\to 0$,
any dependence on $t_i$ (and hence on $m_i$) asymptotically
drops out completely of (\ref{amde}) for $t_L\gg t_i$;
i.e., any memory of the initial condition is
lost for $t\gg t_L\gg t_i$.
In the opposite limit, $w\to\infty$,
the scaling function behaves as
${\cal G}(w)\approx {\cal G}_\infty\,w^{-x_i/z}$,
giving \cite{own}
\begin{equation}
M_{\infty}(L, \,m_i)
\approx \text{const}\times m_i\,L^{z\theta'}\,.
\end{equation}
That is, the amplitude varies linearly in $m_i$,
and its $L$-dependence
is governed by the same exponent $\theta'$
that characterizes the short-time singularity of $m(t)$
in the bulk case.

As we have shown in Ref.\ \onlinecite{own2},
these predictions follow in a natural fashion
from a phenomenological scaling approach.
While they should therefore be
valid for general values of $n$, detailed checks
and explicit calculations of the scaling function
${\cal G}$ have so far been undertaken with only
two methods. One, utilized in Ref.\ \onlinecite{own}, is the
exact solution of the finite-sized $n$-component model A
in the limit $n\to\infty$.
This yielded for the amplitude in (\ref{amde})
the simple form
\begin{equation}\label{minfn}
M_\infty^{(n=\infty)}=
{A\,m_it_L^{\epsilon/4}\over \sqrt{1+B\,m_i^2t_L}}\;,
\end{equation}
where $A$ and $B$ are nonuniversal constants.

The second method used so far is Monte Carlo simulation
of two and three-dimensional Ising models \cite{mont,gras}.
Both Li et al.\ \cite{mont} and
Grassberger \cite{gras} estimated the short-time exponent
$\theta'$, obtaining values in reasonable agreement with
the $\epsilon$-expansion results
of Janssen et al.\ \onlinecite{jans}.
In Ref.\ \onlinecite{mont}, also the dependence of $M_{\infty}$
on $m_i$ was investigated and found to conform remarkably well
with the large-$n$ result (\ref{minfn}).

In the present paper we wish to apply an alternative powerful approach,
one which works for general $n$: the RG approach that uses $\epsilon=4-d$ as
small parameter. From previous studies
\cite{gold1,gold,ZJ,niel,fsli}
it is well known that a proper treatment of finite-size effects
by this method requires a rather {\it subtle} procedure.
Unlike to the case of bulk critical behavior, a dimensionality
expansion in integer powers of $\epsilon$ does not in general
result. The reason is that the
lowest-energy mode of the order-parameter
field $\phi(\bbox{x},t)$, which is the spatially homogeneous one
(with wave number $\bbox{q}=\bbox{0}$) in our case of periodic boundary
conditions, must be treated {\it nonperturbatively}; only the other,
$\bbox{q}\ne\bbox{0}$ modes may be treated by RG-improved perturbation
theory.

A frequently used technique \cite{gold1,niel,fsli,fsBZJ,fsRGJ}
is to integrate out the latter modes in order
to obtain an effective action for the former.
Upon using the functional-integral formulation of our dynamic model,
this could in principle also be done in the present case.
Due to the fact that even the free propagator of the
$\bbox{q}\ne\bbox{0}$ modes involves the $\bbox{q}=\bbox{0}$
part of $\phi$ as a background field,
the resulting effective action becomes
rather complicated, however, and does not lend itself easily to
direct calculations of time-dependent quantities such as $m(t)$.
We have found it more convenient to perform the
perturbative elimination
of the $\bbox{q}\ne\bbox{0}$ modes directly on the level of the
Langevin equation, following Goldschmidt's procedure
in \onlinecite{gold} in this regard. At lowest nontrivial order
of renormalized perturbation theory the resulting
equation for the $\bbox{q}=\bbox{0}$ mode takes the form of
a nonlinear stochastic equation. The required
nonperturbative treatment of
this mode then amounts to solving this equation
by numerical means. Upon averaging the solutions over the
associated noise, the desired $m(t)$ follows.

Our analysis goes in two respects beyond the related one by
Goldschmidt \onlinecite{gold}.
First of all, we take care to properly build in
the initial conditions. This is a central
purpose of the present work. The advantage is twofold.
On the one hand, our approach encompasses
an adequate description of
the anomalous short-time behavior. On the other hand,
it permits us to study in a systematic fashion
how the initial condition $m_i$
manifests itself on long time scales. In this regard, our
work complements --- and in parts corrects ---
Ref.\ \onlinecite{gold}.

Second, we consider the finite-sized model A for general component
number $n$. Accordingly our findings can be compared both with the
exact results for the limit $n\to\infty$ \cite{own,own2}
as well as
with Monte Carlo calculations for the kinetic Ising model
($n=1$) \cite{mont} and for polymers ($n=0$) \cite{gras}.

The remainder of this paper is organized as follows.
In Sect.\ II we first  recall the definition of the finite-sized
$n$-component model A. Starting from the stochastic equation
for the order-parameter field $\phi(\bbox{x},t)$,
we  eliminate the
$\bbox{q}\ne \bbox{0}$ degrees of freedom to obtain
the reduced stochastic dynamics for the $\bbox{q}=\bbox{0}$
component $\Phi(t)$ of $\phi$. Section III deals with the
finite-size scaling forms of the stochastic equation for $\Phi(t)$
and of its solution. We first reconsider briefly the problem
in the case $d>4$. Upon renormalizing the theory for $d\le 4$,
RG equations are then derived for $\Phi(t)$ and its
stochastic equation whose solutions yield the desired
asymptotic scaling forms. Subsequently the so-obtained
scaling form of the stochastic
equation is explicitly verified by means of RG-improved perturbation
theory up to one-loop order within the framework of the resulting
$\sqrt{\epsilon}$ expansion.
In Sect.\ IV we first show that the scaled
stochastic equation at bulk criticality
has asymptotic solutions with the correct
limiting behavior in both the short-time regime as well as in the
long-time one. The scaled stochastic equation is extrapolated to
$\epsilon=1$, and numerical results for its solutions are presented.
These are compared with the large-$n$ results of Ref.\ \onlinecite{own}
and the Monte Carlo simulations of Refs.\ \onlinecite{mont} and
\onlinecite{gras}. Section V contains a brief summary and concluding
remarks. Some technical details have been relegated to two appendices.

\section{Definition of the model and reduced dynamics for the
zero-momentum mode}

We consider the $n$-component model $A$
for a $d$-dimensional hypercube
$V$ with volume $L^d$ under periodic boundary conditions. This is
defined by the Langevin equation  \cite{hoha}
\begin{equation}
\partial_t\, \phi(\bbox{x},t) = - \lambda_0
\frac{\delta {\cal H}\{\phi\}}{\delta \phi(\bbox{x},t)}
+ \zeta (\bbox{x},t)  \label{langevin}
\end{equation}
with the Ginzburg-Landau Hamiltonian
\begin{equation}\label{gila}
{\cal H}\{\phi\} = \int_V \left[\frac{1}{2}\,
(\nabla \phi)^2 +\frac{\tau_0}{2}\,
 \phi^2 + \frac{g_0}{4!n}\, (\phi^2)^2\right]\;,
\end{equation}
where $\phi = (\phi^{\alpha})$ is an $n$-component field, while
$\zeta=(\zeta^{\alpha})$ is a Gaussian random force with mean zero
and correlation
\begin{equation}\label{stoch}
\langle \zeta^{\alpha}(\bbox{x},t)\,\zeta^{\beta}
(\bbox{x}',t')\rangle =2\lambda_0\,\delta^{\alpha \beta}\,\delta
(\bbox{x}-\bbox{x}')\,\delta(t-t') \;.
\end{equation}
All times $t$ and $t'$ are assumed to be positive.
At $t=0$ we impose the initial condition
\begin{equation}\label{inicond}
\phi^\alpha (\bbox{x},0)=\delta^{\alpha 1}\,m_i\;,
\end{equation}
where $m_i$ is independent of $\bbox{x}$.

In view of the chosen periodic boundary conditions
we make a Fourier expansion
\begin{equation}\label{Fourierexp}
\phi^\alpha(\bbox{x},t)=L^{-d}\sum_{\bbox{q}}
\phi^\alpha_{\bbox{q}}(t)\,e^{i\bbox{q}\cdot\bbox{x}}\;,
\end{equation}
where the sum runs over all $\bbox{q}={2\pi\over L}\bbox{m}$ with
$\bbox{m}\in Z\!\!\!Z^d.$
Next
we decompose
\begin{equation}\label{phidecomp}
\phi^{\alpha}({\bbox{x}},t)=\phi^{\alpha}_{\bbox{0}}(t)+
\psi^{\alpha}({\bbox{x}},t)
\end{equation}
into its  $\bbox{q=0}$ part
\begin{equation}
\phi^\alpha_{\bbox{0}}(t)= L^{-d}\int_V
\phi^\alpha(\bbox{x},t)\;,
\end{equation}
and
a remainder $\psi=(\psi^\alpha)$
given by the sum of all ${\bbox{q}}\ne {\bbox{0}}$
terms on the right-hand side of
(\ref{Fourierexp}). An analogous decomposition
we make for the noise, writing
\begin{equation}\label{nodecomp}
\zeta({\bbox{x}},t)=\Xi(t)+\vartheta({\bbox{x}},t)\;.
\end{equation}
As a consequence of (\ref{stoch}) the ${\bbox{q}}={\bbox{0}}$ part
\begin{equation}
\Xi(t)\equiv L^{-d}\int_V\zeta(\bbox{x},t)
\end{equation}
is a Gaussian white noise with correlation
\begin{equation}
\left\langle\Xi^\alpha(t)\,\Xi^\beta(t')\right\rangle
=2\lambda_0\, L^{-d}\,\delta^{\alpha\beta}\,\delta(t-t')\;.
\end{equation}
The $\bbox{q}\ne\bbox{0}$ parts $\psi$ and $\vartheta$ satisfy
\begin{equation}\label{intvartheta}
\int_V\psi(\bbox{x},t)=0\;.
\end{equation}

By averaging (\ref{langevin}) over the ${\bbox{q}}\ne{\bbox{0}}$ part
$\vartheta$ of the noise we can formally obtain a
stochastic equation for the constant mode alone.
Let us denote
an average over $\vartheta$ by $\langle .\rangle_\vartheta$.
Further, we introduce the notation
\begin{equation}\label{decomp2}
\phi^{\alpha}_{\bbox{0}}\equiv \Phi^{\alpha}(t)+\delta\phi^{\alpha}(t)\>,
\end{equation}
with $\langle \delta\phi^{\alpha}(t) \rangle_\vartheta=0$.
Because of (\ref{intvartheta}) and translational invariance
we also have
$\langle\psi(\bbox{x},t)\rangle_\vartheta=0$.
The resulting stochastic equation for $\Phi(t)$ can be written in the form
\begin{equation}\label{exstocheq}
\dot \Phi^{\alpha}(t)+ \lambda_0 \tau_0 \Phi^{\alpha}(t)
+\frac{\lambda_0 g_0}{6}
S^{\alpha\beta\gamma\delta}
\left[\Phi^{\beta}(t) \Phi^{\gamma}(t) \Phi^{\delta}(t)
+ 3 \Phi^{\beta}(t)\,
{\cal C}^{\gamma\delta}(t)+{\cal C}^{\beta\gamma\delta}(t)\right]
= \Xi^{\alpha}(t)\;,
\end{equation}
where
\begin{equation}\label{Stensor}
S^{\alpha\beta\gamma\delta}\equiv
\frac{1}{3}\left(\delta^{\alpha\beta}\,
\delta^{\gamma\delta} + \delta^{\alpha\gamma}\,
\delta^{\beta\delta} +
\delta^{\alpha\delta}
\,\delta^{\beta\gamma}\right)
\end{equation}
and repeated indices are summed over.
A systematic derivation of Eqn.\,(\ref{exstocheq}) is given in Appendix A.
The functions
${\cal C}^{\alpha\beta}(t)$ and
${\cal C}^{\alpha\beta\gamma}(t)$ in (\ref{Stensor})
are the expectation values of bilinear and trilinear
products of $\psi$'s and $\delta\phi$'s.
These are independent of $\bbox{x}$ but depend, of course,
in a complicated manner on $\Phi(t')$ with $0\le t'<t$.
In terms of $\Phi$ the initial condition
(\ref{inicond}) becomes
\begin{equation}\label{inicondphi}
\Phi^\alpha(0)=\delta^{\alpha 1}\,m_i\, ,
\end{equation}
and expectation values involving $\psi$'s and $\delta\phi$'s have
to satisfy Dirichlet initial conditions.

Eqn.\,(\ref{Stensor}), which is an exact consequence of
(\ref{langevin})--(\ref{stoch}), forms the basis of our subsequent
analysis. We emphasize that the functions
${\cal C}^{\alpha\beta}$ and
${\cal C}^{\alpha\beta\gamma}$ are functionals of $\Phi$. In order for
(\ref{langevin}) to become a closed equation for $\Phi$,
this functional dependence must be known. This is the case ---
at least in principle --- if we
consider these functions to be given by their Feynman graph
series. Explicit expressions for their graphs to two-loop order
are presented in Appendix A. The $\Phi$-dependent free propagators
involved --- and hence the resulting expressions for the graphs ---
differ from the ones used by Goldschmidt \cite{gold}. This is
because we imposed an initial condition at $t=0$ rather than in
the infinite past.

\section{Scaling form of stochastic equation and of its solution}

We are interested in the asymptotic solutions of the stochastic
equation (\ref{exstocheq}) near the bulk critical point. Since
these are known to have a scaling form, the stochastic equation
itself should take a scaling form on sufficiently large length and
time scales. In order to show this and to compute explicit
expressions for these scaling forms, we will largely
follow Ref.\ \onlinecite{gold}.

Differences arise because of our imposed initial condition
at $t=0$. In the case $d<4$, essential modifications are
required to ensure a proper treatment of both the short-time
anomalies and the long-time behavior. In the case  $d>4$,
the differences are marginal as far as the leading asymptotic
behavior is concerned,
and the results could be gleaned from Ref.\ \onlinecite{gold}.
Since the subsequent analysis for $d<4$ makes substantial
use of these, it is useful to recapitulate them briefly.

\subsection{The case $d>4$}

Above the upper critical dimension $d^*=4$, the effect of the
$\bbox{q}\ne\bbox{0}$ modes (i.e.\ of $\psi$) is asymptotically
negligible. Thus
${\cal C}^{\alpha\beta}$ and ${\cal C}^{\alpha\beta \gamma}$
may be replaced by zero, their zero-loop
expression,  in (\ref{exstocheq}). Inspection of the resulting stochastic
equation
reveals that it takes a scaling form. In summary, it follows
that $\Phi$ asymptotically behaves as
\begin{equation}\label{scphidg4}
\Phi(t)\approx g_0^{-1/4} L^{-d/4}\,{\hat\Phi}\!\!
\left({g_0^{1/2}\lambda_0t\over L^{d/2}},
{\tau_0L^{d/2}\over g_0^{1/2}},g_0^{1/4} L^{d/4}m_i;
\left\{{L^{3d/4}\,\Xi\over \lambda_0g_0^{1/4}}\right\}\right)\; ,
\end{equation}
where the scaling function ${\hat\Phi}=({\hat\Phi}^\alpha)$
is a solution to
the scaled stochastic equation
\begin{equation}
\left[\partial_{\sf t}+1+\frac16\,
|{\hat\Phi}({\sf t})|^2\right]\!{\hat\Phi}^\alpha({\sf t})
=\hat\Xi^\alpha({\sf t})
\end{equation}
with the initial condition
\begin{equation}
{\hat\Phi}^\alpha({\sf t}=0)=\delta^{\alpha 1}g_0^{1/4} L^{d/4}m_i\;.
\end{equation}
Here ${\sf t}\equiv g_0^{1/2}L^{-d/2}\lambda_0\,t$ is the scaled time,
and the scaled noise
$\hat\Xi({\sf t})\equiv \lambda_0^{-1}g_0^{-1/4} L^{3d/4}\, \Xi(t)$
obeys
\begin{equation}\label{2momxi}
\left\langle\hat\Xi^\alpha({\sf t})\,\hat
\Xi^\beta({\sf t}')\right\rangle=
2\delta^{\alpha\beta}\,\delta({\sf t}-{\sf t}')\;.
\end{equation}

The result (\ref{scphidg4}) is a manifestation of
the familiar fact that finite size scaling \cite{meffss,suzuki}
gets modified above the upper critical dimension
\cite{Brezin,BinderYoung,gold,ZJ,fsli}.
Below it the finite size scaling forms involve the
temperature field $\tau\sim(T-T_c)/T_c$ and the time $t$ in
the combinations $\xi/L$ and $t/L^z$, respectively,
where $\xi\sim\tau^{-\nu}$
is the equilibrium bulk correlation length. This simple form
of finite size scaling breaks down for $d>4$ because $g_0$ gives
rise to a further length that must be taken into account in the
scaling theory.

A second and trivial remark concerns the scaling of $m_i$.
Since the scaling form (\ref{scphidg4}) is dictated by
dimensional considerations, $m_i$ necessary scales with
$L$ in precisely the same fashion as $\Phi$ and hence $m(t)$.
Only for $d<4$ can the corresponding scaling dimensions
$x_i$ and $x_\phi$ become different.

Finally, we note that corrections to the leading asymptotic term
given on the right-hand side of (\ref{scphidg4}) can be computed
by perturbation theory in $g_0$, provided the ultraviolet
singularities have been regularized.

\subsection{The case $d<4$}

\subsubsection{Renormalization}

In this case the influence of the $\bbox{q}\ne\bbox{0}$ part
$\psi$ on $\Phi$ must {\it not} be neglected. Since perturbation theory
in $g_0$ is known to break down in the critical regime, it
cannot be used to treat this influence.
We will rely on RG-improved perturbation theory near $d=4$.
To this end we must first renormalize the theory.
Utilizing the scheme of dimensional regularization
and minimal subtraction of poles in $\epsilon$, we
introduce renormalized quantities through
\begin{equation}\label{phiren}
\phi(\bbox{x},t)=Z_\phi^{1/2}\,\phi_R(\bbox{x},t)\;,\quad
\tilde\phi(\bbox{x},t)=Z_{\tilde\phi}^{1/2}\,\tilde\phi_R(\bbox{x},t)\;,
\end{equation}
\begin{equation}\label{tauren}
\tau_0-\tau_{0,c}=\mu^2 Z_\tau \tau\;,
\quad g_0K_d=\mu^\epsilon Z_uu\;,\quad
\lambda_0=\mu^{-2}\left(Z_\phi/Z_{\tilde{\phi}}\right)^{1/2}\lambda\;,
\end{equation}
\begin{equation}\label{dotphii}
\left.\dot{\phi}\right|_{t=0}=\left(Z_{\phi}Z_i\right)^{1/2}
\left[\left.\dot\phi \right|_{t=0}\right]_R\,,\quad
\left.\tilde{\phi}\right|_{t=0}=\left(Z_{\tilde{\phi}}Z_i\right)^{1/2}
\left[\left.\tilde{\phi} \right|_{t=0}\right]_R\,,
\end{equation}
and
\begin{equation}\label{zetaren}
\zeta(\bbox{x},t)=\left(Z_\phi/
Z_{\tilde\phi}\right)^{1/4}\,\zeta_R(\bbox{x},t)\;.
\end{equation}

Here $\tilde\phi$ is the auxiliary field that is needed in the
functional-integral
formulation of the dynamics
specified by (\ref{langevin})--(\ref{stoch}). We have absorbed
the usual angular factor $K_d=2/[(4\pi)^{d/2}\Gamma(d/2)]$
in $u$.
The momentum scale $\mu$ is arbitrary and
will be set to $\mu=1$ in the sequel
whenever it is suppressed.  Further, $Z_\phi$, $Z_{\tilde\phi}$,
$Z_\tau$, and $Z_u$ are standard renormalization factors.
Finally, $Z_i$ is the analog of the renormalization
factor denoted $Z_0$ by Janssen et al.\ \cite{jans}.

For the associated RG functions we use the notation
\begin{equation}
\eta_\kappa(u)\equiv\left.\mu\partial_\mu\right|_0 \ln Z_\kappa\;,
\quad\kappa=\phi,\,\tilde\phi,\,\tau,\,u,\,i\;,
\end{equation}
and
\begin{equation}\label{etas}
\beta_u\equiv\left.\mu\partial_\mu\right|_0 u
=u[-\epsilon +\eta_u(u)]\;,
\end{equation}
where $\left.\partial_\mu\right|_0$ means a derivative
at fixed bare variables. Using familiar RG arguments,
the critical exponents can be expressed in terms of the
values $\eta_\kappa^*$ of the
functions (\ref{etas}) at the infrared-stable
zero $u^*$ of $\beta_u$.  Specifically for the scaling dimensions
$x_\phi$ and $x_i$ and the critical exponents $z$ and $\theta'$ one has
\cite{janszep}
\begin{equation}
x_\phi={d-2+\eta_\phi^*\over 2}\;,\quad
x_i=x_\phi-{\eta_\phi^*+\eta_{\tilde\phi}^*+\eta_i^*\over 2}\;,\quad
z=2+{\eta_{\tilde\phi}^*-\eta_\phi^*\over 2}\;,\quad
\theta'=-{\eta_\phi^*+\eta_{\tilde\phi}^*+\eta_i^*\over 2z} \;.
\end{equation}
In the one-loop calculation presented below
we shall need the ${\cal O}(u)$ results
\begin{equation}\label{Ztaug}
Z_\tau=1+{n+2\over 6\epsilon}\,u+{\cal O}(u^2)\;,\quad
Z_u=1+{n+8\over 6\epsilon}\,u+{\cal O}(u^2)\;,
\end{equation}
\begin{equation}\label{Zphiphit}
Z_\phi=1+{\cal O}(u^2)\;,\quad Z_{\tilde\phi}=1+{\cal O}(u^2)\;,
\end{equation}
and the $\epsilon$ expansions
\begin{equation}
u^*={6\epsilon\over n+8}+ {\cal O}(\epsilon^2)\;
\end{equation}
and
\begin{equation}\label{thetaepsex}
\theta'={n+2\over n+8}\,{\epsilon\over 4}
+{\cal O}(\epsilon^2)\;.
\end{equation}

We must also discuss what happens to the initial condition
(\ref{inicondphi}) upon renormalization.
This condition can be incorporated
into the stochastic equation (\ref{exstocheq}) by adding
the term $\delta (t)\,m_i\Phi^1(t)$ on the right-hand side.
In the equivalent functional-integral representation of the theory,
such a term corresponds to a contribution of the form
$-\int_Vm_i\,\tilde\phi^1(\bbox{x},0)$ to the action
(cf.\ Eq.\ (5.28) of Ref.\ \onlinecite{jans}). The fact
that the renormalization (\ref{dotphii}) of
$\left.\tilde\phi\right|_{t=0}$ involves an
extra factor $Z_i^{1/2}$ implies that $m_i$ should be reparametrized
as (see Eq.\ (67) of Ref.\ \onlinecite{janszep})
\begin{equation}\label{miR}
m_i=\left(Z_{\tilde\phi} Z_i\right)^{-1/2}m_{i,R}\;.
\end{equation}

\subsubsection{RG equations and finite-size scaling forms}

In this part, we will only be concerned with the renormalized fields
$\Phi_R$ and $\Xi_R$, but not with the bare ones. To simplify the
notation, we shall drop the subscript $R$ on these
as well as on $m_{i,R}$. Thus $\Phi$, $\Xi$, and $m_i$ now
mean renormalized quantities.

By exploiting the arbitrariness of $\mu$ in the above reparametrization
relations, RG equations for $\Phi$ and
its stochastic equation can
be derived in a standard manner.
Let us write the (renormalized) stochastic equation for $\Phi$ as
\begin{equation}
{\cal K}\left(t,L,\tau,u,\lambda,\mu,\{\Phi,\Xi\}\right)=0
\end{equation}
and introduce the operator
\begin{equation}
{\cal D}_\mu=\mu\partial_\mu+\beta_u\partial_u-
\left(2+\eta_\tau\right)\tau\partial_\tau+
\frac12\left(4+ \eta_{\tilde\phi}-\eta_\phi
\right)\!\lambda\partial_\lambda\;.
\end{equation}
Then the RG equations for $\Phi=(\Phi^\alpha)$ and
${\cal K}=({\cal K}^\alpha)$
can be written as
\begin{equation}
\left[{\cal D}_\mu+
{\eta_\phi\over 2}+{\eta_{\tilde\phi} +\eta_i\over 2}\,
m_i\frac{\partial}{\partial m_i}
-{\eta_\phi-\eta_{\tilde\phi}\over 4}
\int\limits_0^\infty \!dt\,
\Xi^\alpha(t){\delta \over\delta\Xi^\alpha(t)}
\right]\!\Phi=0
\end{equation}
and
\begin{equation}
\left[{\cal D}_\mu-
{\eta_{\tilde\phi}\over 2}
-\int\limits_0^\infty \!dt\left(
{\eta_\phi-\eta_{\tilde\phi}\over 4}\,
\Xi^\alpha(t){\delta \over\delta\Xi^\alpha(t)}
+{\eta_{\phi}\over 2}\,
\Phi^\alpha(t){\delta \over\delta\Phi^\alpha(t)}\right)
\right]\!{\cal K}=0\;.
\end{equation}

We solve these by characteristics. The solutions involve
metric factors, given by integrals of exponentials
of $\eta_\kappa$ functions along RG-trajectories. To account for
the nontrivial $\epsilon$-dependence of the scales implied by
the $g_0$-dependence of the zero-loop
result (\ref{scfPhi}), it is convenient to split off appropriate
powers of a constant $G_*\sim \epsilon$ from the metric factors.
We leave open the choice of $G_*$ at this point; the choice
(\ref{Gstar}) we will finally make is such that the effective
coupling constant appearing in the scaled form of the stochastic
equation of motion when $T=T_c$ approaches one in the long-time limit.

The resulting asymptotic
finite-size scaling forms then become
\begin{equation}\label{scfPhi}
\Phi(t,L,m_i,\tau,u,\lambda,\{\Xi\})
\approx G_*^{-1/4} {\cal B}_\phi\, L^{-x_\phi}\,\hat\Phi\!
\left({\sf t},{\sf t}_i,y,
\left\{\hat\Xi\right\}
\right)
\end{equation}
and
\begin{equation}\label{scfeqm}
{\cal K}\!\left(t,L,\tau,u,\lambda,\mu,\{\Phi,\Xi\}\right)
\approx \lambda G_*^{1/4} {\cal B}_\phi^{-1} L^{-\Delta/\nu}\,
\hat{\cal K}\!\left({\sf t},y,
\left\{\hat\Phi,\hat\Xi\right\}\right)\;,
\end{equation}
with  the familiar
magnetic scaling index $\Delta/\nu=d-x_\phi$ and the
scaled arguments
\begin{equation}
{\sf t}=t\lambda G_*^{1/2}{\cal B}_\lambda  L^{-z}\equiv t/t_L\;,
\end{equation}
\begin{equation}
{\sf t}_i=\left({\cal B}_i m_i\right)^{-z/x_i}
G_*^{1/2}{\cal B}_\lambda  L^{-z}
\equiv t_i/t_L\;,
\end{equation}
\begin{equation}
y=G_*^{-1/2}{\cal B}_\tau\tau L^{1/\nu}\;,
\end{equation}
and
\begin{equation}
\hat\Xi=\lambda^{-1}G_*^{-1/4}{\cal B}_\lambda^{-1/2}
L^{(d+z)/2}\,\Xi\;,
\end{equation}
where ${\cal B}_\phi={\cal B}_\phi(u)$,
${\cal B}_\tau$, ${\cal B}_\lambda$, and ${\cal B}_i$ are
nonuniversal metric factors. Consistency requires that the scaled
variable $\hat\Phi$ in (\ref{scfeqm}) be given by
\begin{equation}\label{phihatcon}
\hat\Phi=G_*^{1/4}{\cal B}_\phi^{-1}
L^{x_\phi}\,\Phi\;.
\end{equation}
For the second moment of the scaled
(renormalized) noise $\hat\Xi$ we recover (\ref{2momxi}).
The scaling functions $\hat\Phi$ and $\hat{\cal K}$
are given by the functions $\Phi$ and ${\cal K}$
with $\lambda=L=\mu=1$ and $u=u^*$, i.e.,
\begin{equation}\label{scfg*}
\hat{\cal K}({\sf t},y,\{\hat\Phi,\hat\Xi\})=
G_*^{-1/4}\,{\cal K}(G_*^{-1/2}{\sf t},1,\tau=G_*^{1/2}y,u^*,1,1,
\{G_*^{-1/4}\hat\Phi,G_*^{1/4}\hat\Xi\})\;.
\end{equation}

Below we shall explicitly verify the scaling forms (\ref{scfPhi})
and (\ref{scfeqm}) by means of RG-improved perturbation theory
to one-loop order. This will be done in two steps. First,
we shall compute the
scaled equation of motion $\hat{\cal K}$ for arbitrary $y\ge 0$.
Then this will be solved numerically at the critical temperature
to determine the scaling function $\hat\Phi$ for $y=0$.

\subsubsection{Explicit calculation of the scaling form %
of the stochastic equation}

Our starting point is the stochastic equation
(\ref{exstocheq}) for the bare field $\Phi$. (Since
the bare and
renormalized fields $\Phi$ and $\Xi$ agree
at the one-loop
order we will be working, there is no
need to restore the subscript $R$ on the renormalized ones.
Whether $\Phi$ and $\Xi$ mean bare or renormalized fields
should be clear from the context anyway.)

In order to be able to apply RG-improved perturbation theory,
we first must express all bare quantities in terms of
renormalized ones. We begin by substituting the loop expansions
of the functions ${\cal C}^{\alpha\beta}$
and ${\cal C}^{\alpha\beta\gamma}$
discussed in Appendix A into (\ref{exstocheq}),
where we retain only contributions up to one-loop order.
According to this appendix, in which the explicit expressions
for the Feynman graphs of ${\cal C}^{\alpha\beta}$ and
${\cal C}^{\alpha\beta\gamma}$ are given to two-loop order, the latter
function vanishes at one-loop order and hence can be dropped.
The one-loop result for the former
reads in matrix form
\begin{eqnarray}\label{ceij}
\bbox{\cal C}(t)&=&2\lambda_0\,L^{-d}
\int_0^tdt'
\left[A^d\!
\left(
{8\pi^2\lambda_0 L^{-2}t'}
\right)
-1\right]e^{-2\lambda_0\tau_0t'}\,\bbox{\cal E}(t',t)\;,
\end{eqnarray}
with
\begin{equation}\label{nota}
A(x)\equiv\sum_{k=-\infty}^{\infty}e^{-k^2 x}
\end{equation}
and
\begin{equation}
\bbox{\cal E}(t',t)\equiv
\exp\!\left[-\frac13 \lambda_0g_0\int_0^{t'}\!dt''\,
\bbox{\cal M}(t-t'')\right]\,,
\end{equation}
where the
matrix $\bbox{\cal M}(t)$ is defined by
\begin{equation}\label{matrixM}
{\cal M}^{\alpha\beta}(t)
\equiv \Phi^2(t)\,\delta^{\alpha\beta} +
2\Phi^{\alpha}(t)\,\Phi^{\beta}(t)\;.
\end{equation}

Note
that the $t'$ integration in (\ref{ceij})
ranges from $0$ to $t$. In the analogous Eq.\ (3.24) of Ref.\
\onlinecite{gold} this integration starts from $-\infty$.
This is the reason why the anomalous short-time behavior was missed.

A second observation is that the one-loop term (\ref{ceij}) contains
an ultraviolet singularity. To see this one should recall
from the analyses  \cite{fsBZJ,fsRGJ,zinn} of the static theory
that the function $A(x)$ varies as
$
A(x)\approx ({\pi/x})^{1/2}
$ as $x\to 0$. Thus the integrand in (\ref{ceij})
behaves as $t'^{-2+\epsilon/2}$ near zero,
so that the $t'$ integration
diverges at its lower limit.

In order to extract the associated pole parts and
to show that they have the required local form,
we proceed as follows. Suppose we expand
$\bbox{\cal E}(t',t)$ in powers of $t'$. Then only
the terms to linear order in $t'$ will contribute to
the pole parts. Accordingly we subtract and add
in (\ref{ceij}) a term
corresponding to this Taylor expansion to order $t'$.
The analog of $\tilde{\bbox{\cal C}}(t)$ of
$\bbox{\cal C}(t)$ defined through replacement of
$\bbox{\cal E}(t',t)$ by
\begin{equation}\label{Etilde}
\tilde{\bbox{\cal E}}(t',t)=\bbox{\cal E}(t',t)-\openone+
\frac13 \lambda_0\tau_0\,t'\bbox{\cal M}(t)
\end{equation}
in (\ref{ceij}) is regular in $\epsilon$.
The added terms suggest the definition of the
time-dependent (and $L$-dependent)
variables
\begin{equation}\label{tau0t}
\tilde{\tau}_0(t)\equiv \tau_0+\frac{n+2}{3}\,{\lambda_0g_0\over L^d}
\int_0^tdt'\,
{A^d\!
\left(
{8\pi^2\lambda_0 L^{-2}t'}
\right)
-1\over\exp\left(2\lambda_0\tau_0t'\right)}
\end{equation}
and
\begin{equation}\label{g0t}
\tilde{g}_0(t)\equiv g_0-\frac{n+8}{3}\,{2\lambda_0^2g_0^2\over L^d}
\int_0^tdt'\,t'\,
{A^d\!
\left(
{8\pi^2\lambda_0 L^{-2}t'}
\right)
-1\over\exp\left(2\lambda_0\tau_0t'\right)}\;.
\end{equation}
In terms of these quantities the stochastic equation
that holds for $\Phi(t)$ at one-loop order can be
written as
\begin{equation}\label{laco2}
\left\{\left[\partial_t + \lambda_0 \tilde{\tau}_0(t)
+\frac16\lambda_0\tilde{g}_0(t)\,\Phi^2(t)\right]\!\delta^{\alpha\beta}
+\frac12 \lambda_0g_0\,S^{\alpha\beta\gamma\delta}
\tilde{\cal C}^{\gamma\delta}(t)\right\}\!\Phi^\beta(t)
=\Xi^\alpha(t)\;.
\end{equation}

The Laurent expansions of the integrals
appearing in (\ref{tau0t}) and (\ref{g0t}) are computed in
Appendix B. The results are given in (\ref{I0})
and (\ref{I1}). We substitute  these
into (\ref{tau0t}) and (\ref{g0t}). Upon
expressing $g_0$, $\tau_0$, and $\lambda_0$ by means of
(\ref{phiren})--(\ref{tauren})
and (\ref{Ztaug}) in terms of their
renormalized analogs, the one-loop poles $\sim u/\epsilon$ in
$\tilde\tau_0(t)$ and $\tilde{g}_0(t)/u$ are found to cancel, giving
\begin{equation}
\tilde{\tau}_0(t)=\tilde\tau(t)+{\cal O}(u^2)\;,\quad
\tilde{g}_0(t)=\tilde{g}(t)+{\cal O}(u^3)\;,
\end{equation}
with the ultraviolet-finite, time-dependent
quantities
\begin{eqnarray}\label{taut}
\tilde \tau(t)&=&\tau+
\frac{n+2}{12}\,u\Bigg\{\tau\left[\ln\tau
-\text{Ei}(-2\lambda \tau t)
-\frac{\exp(-2\lambda t \tau)}{2\lambda\tau t }
 \right]\nonumber\\
& &\mbox{}
 +
{4\over L^2}\, F_0\!\left(8\pi^2\lambda t/L^2,\,
L^2\tau/4\pi^2\right)\! +{\cal O}(\epsilon)\Bigg\}
\end{eqnarray}
and
\begin{eqnarray}\label{gete}
\tilde g(t)&=&K_d^{-1}\Bigg\{u+\frac{(n+8)u^2}{12}\bigg[1+\ln \tau
-\text{Ei}(-2\lambda t \tau)\nonumber\\
& &\mbox{}-
 \frac{1}{\pi^2}\,
F_1\!\left(8\pi^2\lambda t/L^2,\, L^2\tau/4\pi^2
\right)\!\bigg]+{\cal O}(\epsilon)\Bigg\}\;,
\end{eqnarray}
where
$\text{Ei}(x)$ is the exponential-integral function in the notation
of \onlinecite{GradRyz},
and we have introduced the functions
\begin{equation}\label{Fp}
F_p(\alpha,\beta)\equiv
\int_0^{\alpha}\!dx\,x^p\left(A^4(x)-1-\pi^2/x^2\right)\,
{\rm e}^{-\beta x}
\end{equation}
with $p=0,1$. The quantities $\tilde\tau$ and $\tilde g$
are time-dependent analogs of the static shifted variables
introduced in Refs.\ \onlinecite{fsBZJ} and encountered again in
the treatment of the dynamic case in
Ref.\ \onlinecite{fsli} (where a different normalization
of the renormalized coupling constant was employed, however).
In the limit $t\to\infty$, our $t$-dependent variables
$\tilde\tau$ and $\tilde g$ reduce to these,
as they should. That is,
\begin{eqnarray}
\lim_{t\to\infty}\tilde\tau(t)&=&\tilde{\tau}_{\text{st}}
=
\tau+\frac{n+2}{12}\,u\,\tau\!\left[\ln\tau+\frac{4}{L^2}\,
F_0\!\left(\infty,L^2\tau/4\pi^2\right)\right]
\end{eqnarray}
and
\begin{equation}
\lim_{t\to\infty}\tilde{g}(t)=\tilde{g}_{\text{st}}=
K_d^{-1}u\left\{1+\frac{n+8}{12}\,u
\left[1+\ln\tau-\frac{1}{\pi^2}\,
F_1\!\left(\infty,L^2\tau/4\pi^2\right)\right]\right\}.
\end{equation}

In the limit $t\to 0$,
$\tilde\tau(t)$ and $\tilde g(t)$ have short-time singularities
of the form $\tilde\tau(t)\sim t^{-1}+{\cal O}(\tau \ln t)$ and
$\tilde g(t)\sim \ln t$. As we will see below,
the $t^{-1}$ singularity has the correct form and coefficient to
produce the short-time singularity $\sim t^{\theta'}$ of
$\Phi(t)$.

The renormalized stochastic equation
that holds for $\Phi(t)$ at one-loop order
is given by (\ref{laco2}), with
$\tilde \tau_0$, $\tilde g_0$, $g_0$, and
$\lambda_0$ replaced by
$\tilde \tau$, $\tilde g$, $u/K_d$, and
$\lambda$, respectively (and $\Phi$ and $\Xi$ interpreted as
renormalized fields). We now choose the constant $G_*$ in
(\ref{scfPhi})--(\ref{scfg*}) as
\begin{eqnarray}\label{Gstar}
G_*&\equiv& \lim_{\tau\to 0+}\tilde{g}_{\text{st}}(\tau,L=1,u^*)\nonumber\\
&=&K_d^{-1}u^*\left\{1+\frac{n+8}{12}\,u^*
\left[1-C_E+\ln 4\pi^2-\frac{1}{\pi^2}\,f_1\right]\right\}
\end{eqnarray}
with
\begin{equation}
f_1\equiv\int_0^\infty dx\,x\left[A(x)^4-1-{\pi^2 x^{-2}}\,
\theta(1-x)\right],
\end{equation}
where $C_E$ is Euler's constant and $\theta(x)$ denotes the Heaviside
step function.

Upon substitution of the above results
into (\ref{scfg*}) the scaled stochastic equation
$\hat{\cal K}=0$ becomes
\begin{equation}\label{stocheqPhi}
\left[\partial_{\sf t}+\tau_{\text{eff}}(y,{\sf t})+
\frac16\,u_{\text{eff}}(y,{\sf t})\,|\hat\Phi|^2
+{\cal O}\!\left(\epsilon^{3/2}\right)\right]\!\hat\Phi
=\hat\Xi
\end{equation}
with
\begin{eqnarray}\label{taueff}
\tilde{\tau}_{\text{eff}}({\sf t},y)&=&G_*^{-1/2}
\tilde{\tau}^*L^z\nonumber\\
&=&y\left\{1+
\frac{n+2}{12}\,u^*\left[\ln\left(yG_*^{1/2}\right)
-\text{Ei}(-2 y {\sf t})
-\frac{\exp(-2 y {\sf t} )}{2y {\sf t} }
 \right]\right\}\nonumber\\
& &\mbox{}
 +{n+2\over 3}\,\left(K_d u^*\right)^{1/2}\,
F_0\!\left(8\pi^2G_*^{-1/2}{\sf t},\,
G_*^{1/2}y/4\pi^2\right)\! +{\cal O}\!\left(\epsilon^{3/2}\right)
\end{eqnarray}
and
\begin{eqnarray}\label{ueff}
\tilde{u}_{\text{eff}}({\sf t},y)&=&
G_*^{-1}\tilde{g}^*\,L^{z-2x_\phi}\nonumber\\
&=&1+\frac{n+8}{12}\,u^*\bigg\{C_E+\ln\left(G_*^{1/2}y/4\pi^2\right)
-\text{Ei}(-2y{\sf t} )
\nonumber\\& &\mbox{}
+\frac{1}{\pi^2}\left[f_1-
F_1\!\left(8\pi^2G_*^{-1/2}{\sf t},\, G_*^{1/2}y/4\pi^2
\right)\!\right]\bigg\}
+{\cal O}\!\left(\epsilon^2\right).
\end{eqnarray}
The contribution
involving $\tilde{\cal C}^{\alpha\beta}$ would give rise to
a memory term; this
has been omitted since it is of order $\epsilon^{3/2}$.

These equations hold for general values of ${\sf t}$ and $y$.
However, the numerical calculations discussed
in the next section have only been carried out
at the critical point, $y=0$. For this value of $y$ the
expressions (\ref{taueff}) and (\ref{ueff}) simplify to
\begin{eqnarray}
\tilde{\tau}_{\text{eff}}^c({\sf t})\equiv
\tilde{\tau}_{\text{eff}}({\sf t},0)
&=&
{1\over 2\pi\sqrt{3}}\,{n+2\over \sqrt{n+8}}\,\sqrt{\epsilon}\,
F_0\!\left(8\pi^2G_*^{-1/2}{\sf t},0\right)
-\frac{n+2}{n+8}\,\frac{\epsilon}{4}\,\frac{1}{\sf t}
+{\cal O}\!\left(\epsilon^{3/2}\right)
\end{eqnarray}
and
\begin{eqnarray}\label{ueffc}
\tilde{u}_{\text{eff}}^c({\sf t}) \equiv
\tilde{u}_{\text{eff}}({\sf t},0)
&=&1+\frac{\epsilon}{2\pi^2}\int\limits_{8\pi^2G_*^{-1/2}{\sf t}}^\infty\!
dx\,x\left[A(x)^4-1\right]+{\cal O}\!\left(\epsilon^2\right).
\end{eqnarray}
These quantities approach
nonvanishing constants $\tilde{\tau}_{\text{eff}}^c(\infty)$ and
$\tilde{u}_{\text{eff}}^c(\infty)=1$ in the limit ${\sf t}\to \infty$.
In the short-time limit they behave as
\begin{equation}
\tilde{\tau}_{\text{eff}}^c({\sf t}\to 0)\approx -{\theta'\over {\sf t}}
+\sqrt{\epsilon}\,{\cal O}({\sf t}^0)+{\cal O}\!\left(\epsilon^{3/2}\right)
\end{equation}
and
\begin{equation}\label{utildecnef}
\tilde{u}_{\text{eff}}^c({\sf t}\to 0)\approx
1-\frac{\epsilon}{2}\left[
\ln \left(8\pi^2G_*^{-1/2}{\sf t}\right)
-\pi^{-2}f_1+{\cal O}({\sf t})\right]+{\cal O}\!\left(\epsilon^2\right),
\end{equation}
respectively. As we will see below, the
short-time singularity of the latter quantity
should be exponentiated to a power of the form
${\sf t}^{-\epsilon/2+{\cal O}(\epsilon^2)}$.

\section{Numerical solution of scaled stochastic equation}

We now turn to the computation of the finite-size scaling form of the order
parameter
$m(t)$. To this end we must solve the scaled stochastic equation
(\ref{stocheqPhi})
and average the solution $\hat\Phi({\sf t})$ over  the scaled noise
$\hat\Xi({\sf t})$.
Restricting ourselves to the case of relaxation directly at the critical
point, we set
$\tau=y=0$. For notational simplicity, we also put
the time unit $\lambda^{-1}=1$. Further, we fix the nonuniversal
metric factors
${\cal B}_\phi$ and ${\cal B}_\lambda$ such that
$G_*^{-1/4} {\cal B}_\phi=G_*^{1/2}{\cal B}_\lambda=1$, leaving open
the choice of ${\cal B}_i$ for the moment.

Upon averaging the scaling form  (\ref{scfPhi}) of  $\Phi^1(t)$, we see that
at the critical point
$m(t)$ can asymptotically be written as
\begin{equation}\label{scfmc}
m(t,L,m_i)\approx L^{-x_\phi}\,\sigma({\sf t},{\sf t}_i)
={\cal B}_i m_i\,t^{\theta'} f({\sf t},{\sf t}_i)
\end{equation}
with
\begin{equation}
\sigma({\sf t},{\sf t}_i)=
{\sf t}^{\theta'}{\sf t}_i^{-x_i/z}
f({\sf t},{\sf t}_i)=
\left\langle\hat\Phi^1\!
\left({\sf t},{\sf t}_i,0,\{\hat\Xi\}\right)
\right\rangle_{\hat\Xi}
\end{equation}
where $\langle.\rangle_{\hat\Xi}$ means an average over $\hat\Xi$.
The advantage of rewriting the scaling form in terms of $f$ is that
the short-distance singularity $t^{\theta'}$ has been split off.
Hence $\lim_{{\sf t}\to 0}f({\sf t},{\sf t}_i)$ must exist and be
nonvanishing, and since $m(t,L,m_i)/m_i$ must be independent of $m_i$
whenever $t\ll \min (t_L,t_i)$, the limiting value  $f(0,{\sf t}_i)$ must
be independent of ${\sf t}_i$. We choose the normalization
\begin{equation}
f(0,{\sf t}_i)=f(0,0)\equiv 1\,,
\end{equation}
which fixes the nonuniversal constant ${\cal B}_i$.
Upon redefining $m_i$, we absorb this constant in $m_i$,
and hence take ${\cal B}_i=1$ in the sequel.

In general, the solution of the scaled stochastic equation
(\ref{stocheqPhi}) can
only be obtained by numerical integration. However,
some analytical information about the asymptotic behavior in
certain regimes can be extracted. Let us consider (\ref{stocheqPhi})
for small ${\sf t}=t/t_L$. Due to the short-time singularity
of $\tilde\tau_{\text{eff}}^c$, all bounded continuous solutions
of (\ref{stocheqPhi}) must start from zero. (No confusion should arise
from the fact that
$\hat\Phi^1({\sf t},{\sf t_i})$ and hence
the scaling function $\sigma({\sf t},{\sf t_i})$ in (\ref{scfmc})
approach zero as ${\sf t}\to 0$ even though $m_i\ne 0$.
In view of the anticipated short-time singularity
$\sigma\sim {\sf t}^{\theta'}$ and (\ref{theta}) this is seen to be an
immediate consequence of the inequality $x_i>x_\phi$.)

Provided $m_i$ was taken small,
$\hat\Phi^1({\sf t},{\sf t}_i)$
will also be small for $t\ll \min(t_L,t_i)$,
i.e., for ${\sf t}\ll \min(1,{\sf t}_i)$. Hence the nonlinear term may be
neglected in this short-time regime. The solution to the resulting linear
equation (involving the short-time singularity of
$\tilde\tau_{\text{eff}}^c$) is indeed of the form
$\hat\Phi^1({\sf t},{\sf t}_i)\propto {\sf t}^{\theta'}$, with
$\theta'$ given by the ${\cal O}({\epsilon})$ expression
(\ref{thetaepsex}).

As $t$ grows, $\hat\Phi$ grows. Since the
proportionality constant is proportional to $m_i$, this
initial growth is very rapid if $m_i$ is large.
Once $t$ reaches $t_i$,
the nonlinear term in the stochastic equation
(\ref{stocheqPhi}) must no longer be neglected.
The larger $t/t_i$ gets, the more important this term
becomes till it finally dominates completely both the linear term
$\propto \tilde\tau_{\text{eff}}^c$ and the noise. As long as
$t$ remains small in comparison to $ t_L$, we may then use the
asymptotic short-time form
$\tilde u_{\text{eff}}^c\sim {\sf t}%
^{-\epsilon/2+{\cal O}(\epsilon^2)}$.
In this manner one finds that the solution decays as
\begin{equation}
\sigma({\sf t}\ll 1, {\sf t}_i\ll {\sf t})\sim {\sf t}^{-\beta/\nu z}
\end{equation}
in the specified regime,
where the exponent $\beta/\nu z$, which is
known to govern the nonlinear critical relaxation
of the order parameter in bulk systems, takes
the one-loop value $\beta/\nu z=1/2-\epsilon/4$.

In order to determine the asymptotic behavior
in the long-time regime ${\sf t}\gg 1$, one can substitute
$\tilde\tau_{\text{eff}}^c$ and $\tilde u_{\text{eff}}^c$ in (\ref{stocheqPhi})
by their
limiting values $\tilde\tau_{\text{eff}}^c(\infty)$
and $\tilde u_{\text{eff}}^c(\infty)=1$. The resulting stochastic
equation is equivalent to the one investigated  in
previous analyses of finite-size effects on linear relaxation
\cite{gold1,fsli,ZJ}. From these papers and Ref.\ \onlinecite{own}
it is clear that
$\sigma({\sf t},{\sf t}_i)$ decays as
\begin{equation}\label{sigmainfty}
\sigma({\sf t}\to\infty,{\sf t}_i)\approx
\sigma_\infty({\sf t}_i)\;e^{-\text{const}\times{\sf  t}}\;,
\end{equation}
where
the  amplitude
$\sigma_\infty({\sf t}_i)$
according to Ref.\ \onlinecite{own} should depend on ${\sf t}_i$
in a nontrivial fashion (cf.\ the discussion in the
Introduction). In particular, it should have the asymptotic behavior
\begin{equation}\label{asfsigmainf}
\sigma_\infty({\sf t}_i)\approx\left\{\begin{array}{l@{\quad {\rm for}\quad }l}
a&{\sf t}_i\to 0\,,\\
b\,{\sf t}_i^{-x_i/z}=b\,{\sf m}_i&{\sf t}_i\to \infty\,,\end{array}
\right.
\end{equation}
where $a$ and $b$ are nonvanishing universal constants, and
\begin{equation}\label{miscaled}
{\sf m}_i\equiv{\sf t}_i ^{-x_i/z}=m_i\,t_L^{x_i/z}\>.
\end{equation}

We expect that it should be possible to extract analytical information about
$\sigma_\infty({\sf t}_i)$  from the scaled stochastic equation
(\ref{stocheqPhi}).
We have not done this, however. In the following we concentrate
on the numerical solution of (\ref{stocheqPhi}), which provides detailed
information about all regimes discussed above.

For the numerical solution of
(\ref{stocheqPhi}) in $d=3$ dimensions we make the ansatz
\begin{equation}
\hat\Phi^1({\sf t},{\sf t}_i)={\sf t}_i^{-x_i/z}{\sf t}^{\theta'}
f_{\hat\Xi}({\sf t},{\sf t}_i)=m_i\,t_L^{x_i/z}{\sf t}^{\theta'}
f_{\hat\Xi}({\sf t},{\sf t}_i)
\end{equation}
with
\begin{equation}
f_{\hat\Xi}(0,{\sf t}_i)=1\;.
\end{equation}
This leads to the equation
\begin{equation}\label{stocheqf}
\left[\frac{\partial}{\partial {\sf t}}+r_{\text{eff}}^c(\infty)
+\frac16\,\tilde u_{\text{eff}}^c({\sf t})\,|\hat\Phi|^2\right]
f_{\hat\Xi}({\sf t},{\sf t}_i)
={\sf t}_i^{x_i/z}\,{\sf t}^{-\theta'}\hat\Xi^1({\sf t})
\end{equation}
with
\begin{eqnarray}\label{reffc}
r_{\text{eff}}^c({\sf t})&\equiv&
\tilde\tau_{\text{eff}}^c({\sf t})+\frac{\theta'}{\sf t}
=
{n+2\over \sqrt{n+8}}\,\sqrt{\epsilon}\,
F_0\!\left(8\pi^2G_*^{-1/2}{\sf t},0\right)
+{\cal O}\!\left(\epsilon^{3/2}\right)
\end{eqnarray}
and
\begin{equation}
\left|\hat\Phi\right|^2=\left|\hat\Phi^\perp\right|^2+\left(
{\sf m}_i {\sf t}^{\theta'}
f_{\hat\Xi}\right)^2\,,
\end{equation}
where $\hat\Phi^\perp$ denotes the $n-1$ dimensional component of
$\hat \Phi$ in the subspace orthogonal to the $\alpha=1$ direction.

For reasons discussed above it is important that
the short-time singularity of $\tilde u_{\text{eff}}^c({\sf t})$ is
properly exponentiated. We used the  $\epsilon=1$ expression
\begin{equation}\label{ueffcint}
\tilde u_{\text{eff}}^c({\sf t})= 1+  \left[\frac{1}{\sqrt{0.4x}}-\frac12
\right]
e^{-0.6 x}\;,\quad  x=8\pi^2G_*^{-1/2}{\sf t}\;,
\end{equation}
which smoothly interpolates between the asymptotic form
$\tilde u_{\text{eff}}^c\sim{\sf t}^{-\epsilon/2}$ at small ${\sf t}$ and
the large-${\sf t}$ limit  $\tilde u_{\text{eff}}^c(\infty)=1$
(see Fig.\,1).
The constants $0.4$ and $0.6$ in (\ref{ueffcint})
were determined so that the interpolation formula (\ref{ueffcint}) fits the
the numerical results of the expression (\ref{utildecnef})
with $\epsilon=1$ in the large-${\sf t}$ regime .
For $r_{\text{eff}}^c({\sf t})$ we utilized the expression (\ref{reffc}) with
$\epsilon=1$.\\[2mm]

\def\epsfsize#1#2{0.58#1}
\hspace*{0cm}\epsfbox{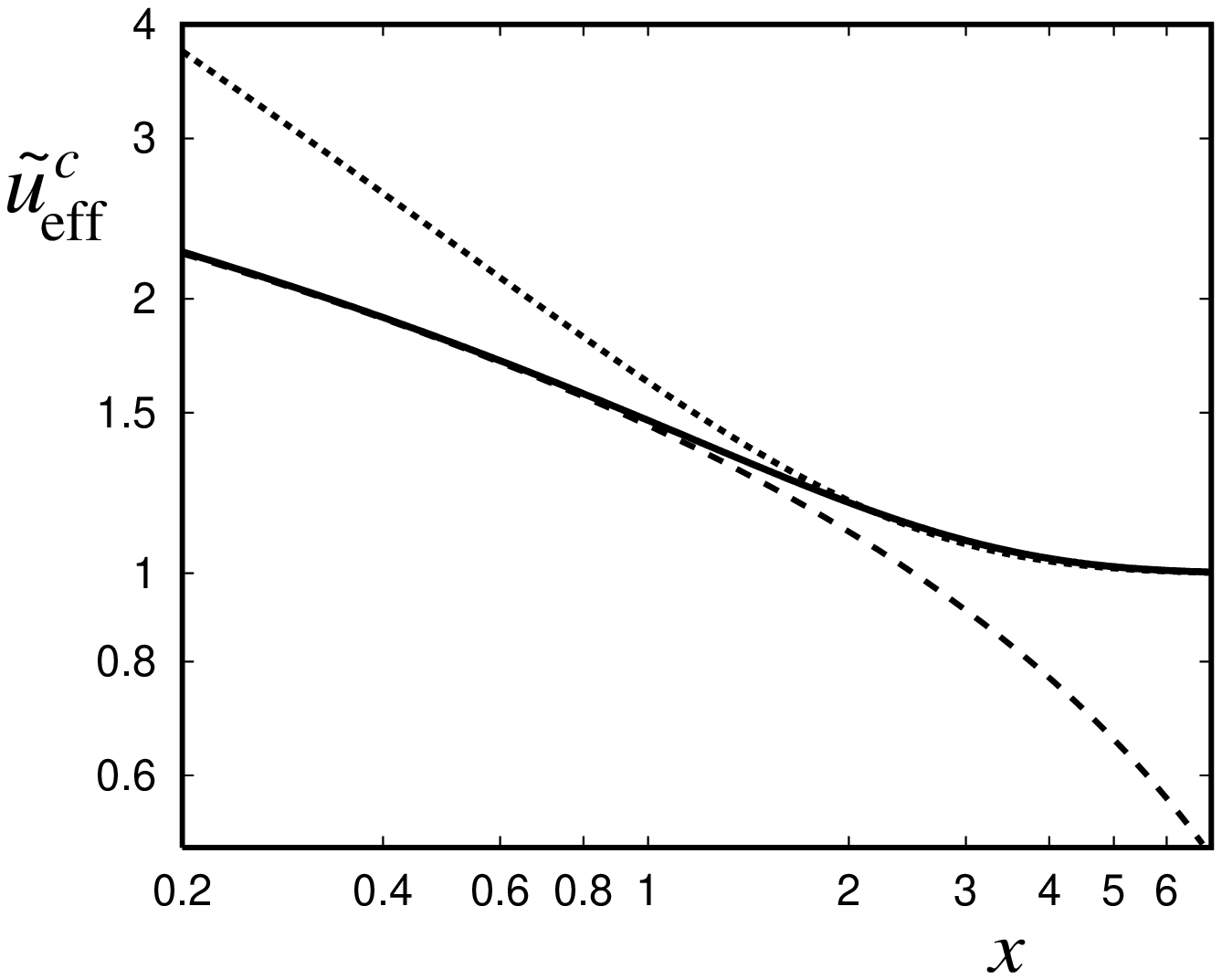}
\vspace*{-6.2cm}\\
\hspace*{9.4cm}
\noindent
\parbox{6.1cm}{{\small {\bf Fig.\,1.:}
$\tilde u_{\text{eff}}^c$ as a function of $x=
8\pi^2G_*^{-1/2} {\sf t}$. The solid line represents the data from
the numerical integration of (\ref{ueffc}) with $\epsilon=1$. The
dashed line shows the asymptotic logarithmic
behavior for $x\to 0$, given in (\ref{utildecnef}). The dotted curve
represents the expression (\ref{ueffcint}), which was used in the numerical
calculation.}}
\\[16mm]

The stochastic equation (\ref{stocheqf}) was
integrated numerically with a first-order Euler procedure
\cite{hohn}. For
$n\neq 1$ one has to solve simultaneously also the original equation
(\ref{stocheqPhi}) for the $n-1$ orthogonal modes. Averages for the
orthogonal modes remain at zero, which means that the order parameter
does not `rotate' within the $n$-dimensional space.
In order to obtain
averages, summations over 30\,000-100\,000
single trajectories were performed.
Exemplary results for $\sigma({\sf t},{\sf t}_i)$
with ${\sf m}_i={\sf t}_i^{-x_i/z}$
as parameter are displayed in semi-logarithmic and
double-logarithmic
representation in Fig.\,2 for $n=1$ and
in Fig.\,3 for $n=20$.\\[5mm]

\def\epsfsize#1#2{0.5#1}
\hspace*{-0.2cm}\epsfbox{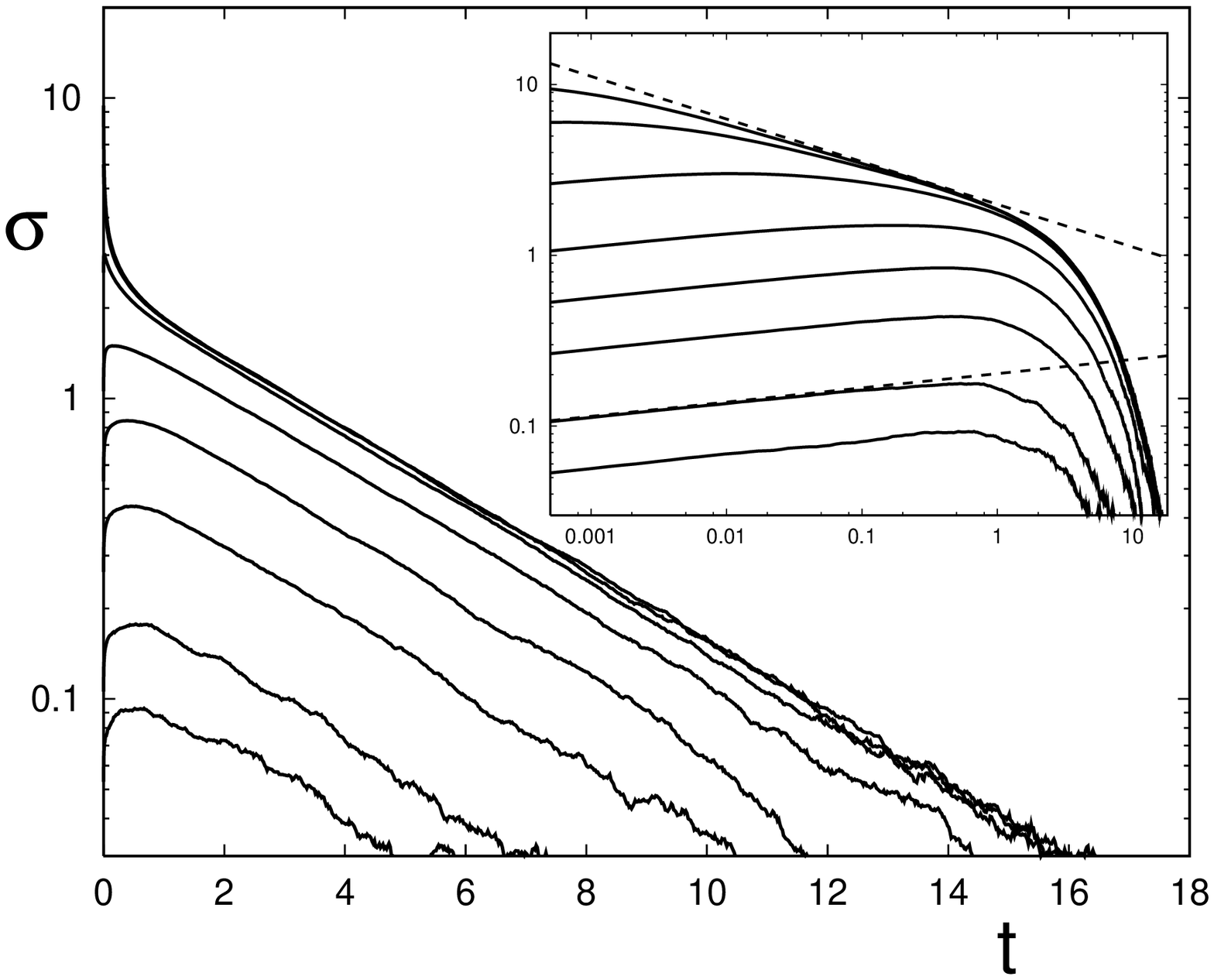}
\vspace*{-7.15cm}\\
\hspace*{9.6cm}
\noindent
\parbox{6.5cm}{{\small {\bf Fig.\,2.:} Order-parameter profiles for
$n=1$ and
${\sf m}_i=0.1,$ $0.2,\,0.5,\, 1,\, 2,$ $5,$ $12,$ $25$ (from bottom to top)
obtained from the numerical solution of (\ref{stocheqf}).
In the main figure the data are displayed in semi-logarithmic form. The
small double-logarithmic diagram inserted shows the same data
emphasizing the power-law growth for small time and the nonlinear
regime for larger values of ${\sf m}_i$ (upper curve). The pure
power laws $\sim {\sf t}^{1/12}$ and
$\sim {\sf t}^{-1/4}$ are represented by the
dashed lines.}}
\\[1.3cm]

\def\epsfsize#1#2{0.48#1}
\hspace*{-0.2cm}\epsfbox{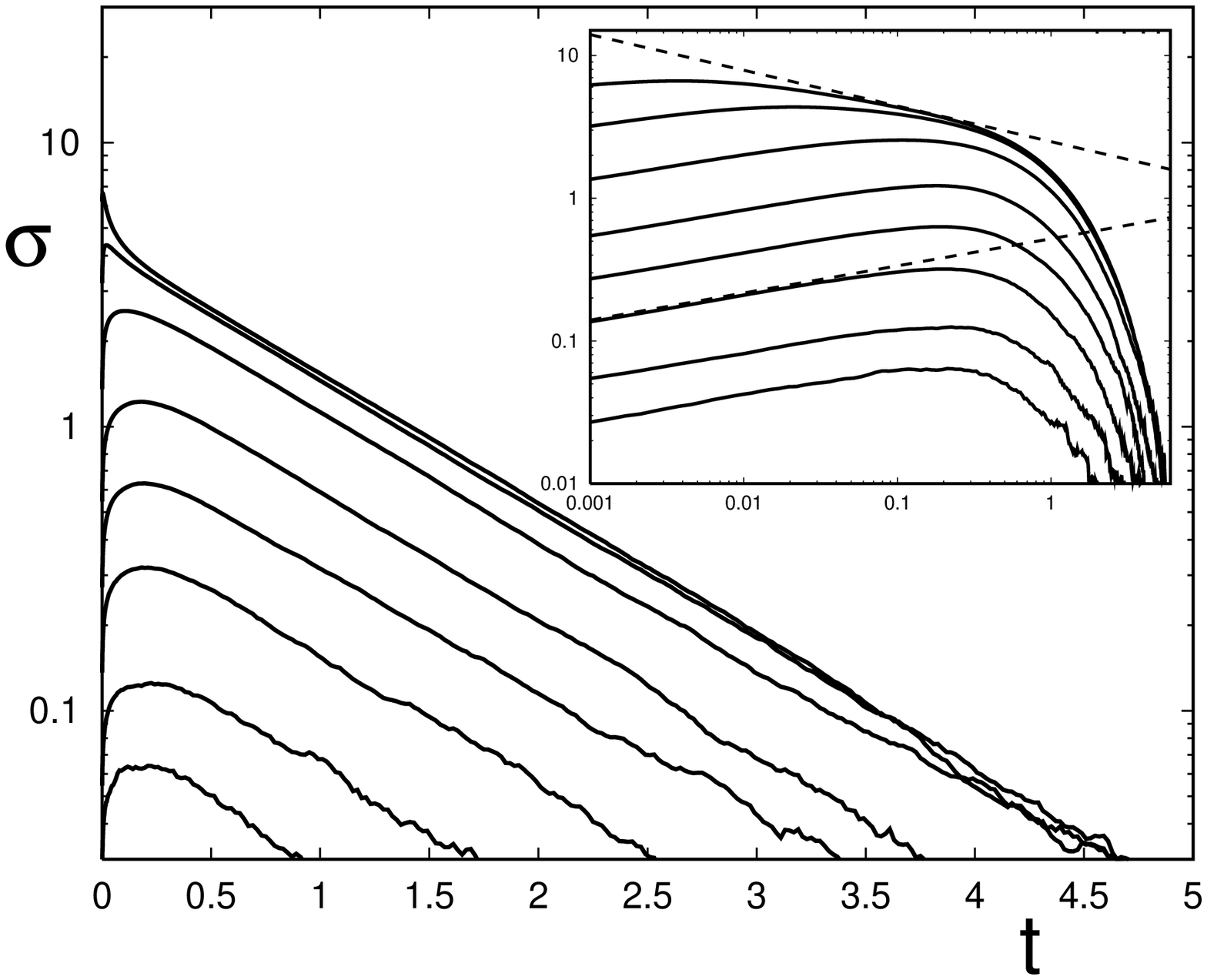}
\vspace*{-7.1cm}\\
\hspace*{9.6cm}
\noindent
\parbox{6.5cm}{{\small {\bf Fig.\,3.:}
Order-parameter profiles for $n=20$ and
${\sf m}_i=0.1,\,0.2,\,0.5,\,1,\,2,$ $5,$ $12,$ $25$ (from bottom to top).
The dashed lines in the small figure show the pure power laws
$\sim {\sf t}^{11/56}$ and $\sim {\sf t}^{-1/4}$.}}
\\[4.9cm]

A common property of all curves
is the expected power-law increase
for small ${\sf t}$. The lower dashed lines in Figs.\,2 and 3
represent the pure power laws ${\sf t}^{1/12}$ and ${\sf t}^{11/56}$, which,
according to (\ref{thetaepsex}),
hold for $n=1$ and $n=20$ in the limit ${\sf t}\to 0$ at the level of our
approximation.
In the large-${\sf t}$ regime
the scaled magnetization decays {\it exponentially} in all cases, with an
${\sf m}_i$-dependent amplitude. For  small ${\sf m}_i$ (bottom
curve), the initial algebraic growth
crosses over directly to the {\it linear} relaxational decay. As
${\sf m}_i$ increases, the
maxima of the curves gradually shift to the left.
Further, immediately to the right of the maxima, a regime
of  {\it nonlinear} relaxational behavior
starts to
emerge, which becomes more and more pronounced as ${\sf m}_i$
grows.

As the results in Figs.\, 2 and 3 show,
the scaling function $\sigma({\sf t},{\sf t}_i)$
becomes {\it independent} of ${\sf m}_i$ in the limit
${\sf m}_i\to\infty$ (${\sf t}_i\to 0$). Its limiting form
$\sigma({\sf t},0)$ has the following properties: At ${\sf t}=0$,
it jumps from $0$ to $\infty$, then it exhibits
nonlinear relaxational decay
$\sim{\sf t}^{-\beta/\nu z}$ until the
crossover to linear relaxational behavior occurs.
This is the special case studied
in the existing literature \cite{gold}.
The top curves in both figures are already
close to this limit.
They
approach the upper dashed lines, which represent the
power law ${\sf t}^{-1/4}$, $1/4$ being the value of
$\beta/\nu z$ in our approximation.
In the opposite
limit
${\sf m}_i\rightarrow 0$ (${\sf t}_i\to \infty$),
the scaling function
$\sigma({\sf t},{\sf t}_i)$
approaches, of course, zero for all ${\sf t}$.\\[2mm]

\def\epsfsize#1#2{0.55#1}
\hspace*{-0.2cm}\epsfbox{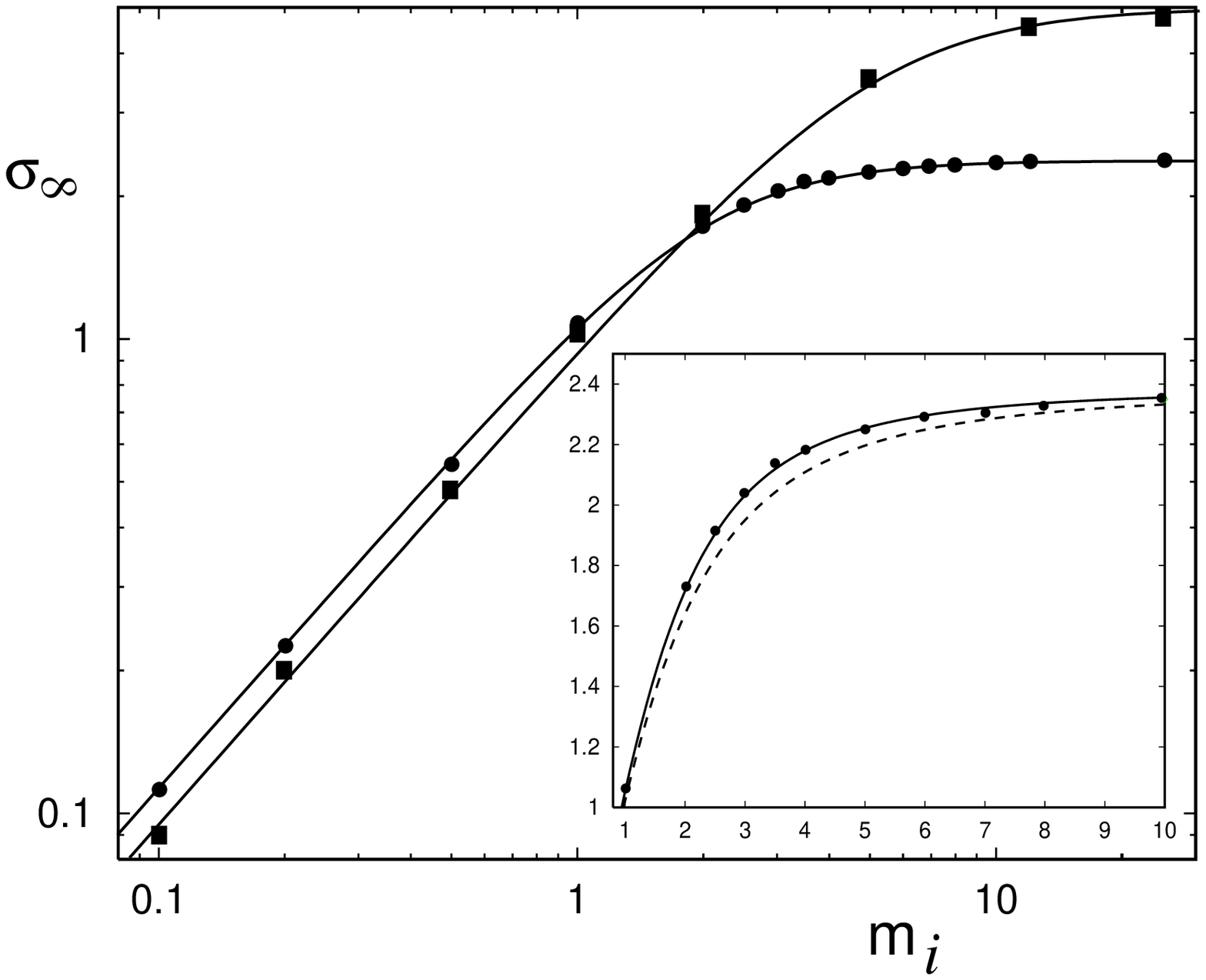}
\vspace*{-7.5cm}\\
\hspace*{10.1cm}
\noindent
\parbox{6.0cm}{{\small {\bf Fig.\,4.:}
The amplitude $\sigma_{\infty}$ defined in
(\ref{sigmainfty}) for $n=1$ (dots) and $n=20$ (squares) as
a function of ${\sf m}_i$
as determined
from the numerical data of Figs. 2 and 3. The curves fitted to the
data are discussed in the text.}}
\\[5.2cm]

We have also determined the amplitude $\sigma_\infty({\sf t}_i)$
defined in (\ref{sigmainfty}). The results
are displayed in Fig.\,4.
They are in conformity with the limiting
forms for large and small ${\sf t}_i$ predicted in (\ref{asfsigmainf}).
It is interesting to compare them with the exact analytical expression
\begin{equation}\label{sigmainfninf}
\sigma_\infty^{(n=\infty)}({\sf t}_i={\sf m}_i^{-z/x_i})=
\frac{a{\sf m}_i}{\sqrt{\left(a/b\right)^2+
{\sf m}_i^2}}
\end{equation}
that holds in the limit $n\to\infty$ \cite{own}.
To this end we have tried to fit
our numerical data to the form (\ref{sigmainfninf}).
For $n=20$ the function
(\ref{sigmainfninf}) provides a good description of the
numerical results (represented by the squares).
For the curve shown in Fig.\,4 the
parameter values are $a=4.92$ and $b=0.96$.
For $n=1$, however, (\ref{sigmainfninf}) yields a
relatively poor fit, especially
in the crossover region. The numerical data
are represented by the full circles. The crossover region is
magnified in the small diagram inserted in Fig.\,4., where
the dashed curve shows the fit obtained with (\ref{sigmainfninf}).
A much better fit is provided by the function
$a\,{\sf m}_i/\left((a/b)^{2.3}+{\sf m}_i^{2.3}\right)^{1/2.3}$,
with $a=2.38$ and $b=1.13$, depicted by the solid curve.

In our discussion of the short-time anomalies of critical relaxation
in the Introduction we pointed out that this phenomenon is intimately
related to the fact that the correlation length
$\xi(t)$ is initially small and grows as time proceeds,
approaching its equilibrium value
$\xi_{\text{eq}}=\xi(\infty)$.
Rather than the correlation length $\xi(t)$,
we preferred to investigate the variances
\begin{equation}\label{vari}
m^{(2)}(t)=\left\langle \Phi^1(t)\,\Phi^1( t)\right\rangle
-m(t)^2\;,
\end{equation}
and
\begin{equation}\label{varitrans}
m^{(2)}_{\perp}(t)=\left\langle \Phi^{\perp}(t)\,\Phi^{\perp}(
t)\right\rangle\;,
\end{equation}
where the latter expression is defined for $n\neq 1$.
At the one-loop level there are no
contributions from the $\bbox{q}\neq\bbox{0}$ noise
to (\ref{vari}) and (\ref{varitrans}). Hence, we have simply
subsituted $\phi_{\bbox{0}}\to \Phi$ in these expressions
(see (\ref{decomp2}) and Appendix A).
The variances (\ref{vari}) and (\ref{varitrans})
are numerically somewhat easier accessible, yet
display a qualitatively similar time dependence as $\xi(t)$.
Restricting ourselves again to the critical case $\tau=0$, we
have the scaling forms
\begin{equation}\label{upsi}
m^{(2)}(t,L,m_i)\approx L^{2-\eta}\,
\Upsilon\left({\sf t},{\sf t}_i\right)\;
\end{equation}
and
\begin{equation}\label{upsitrans}
m_{\perp}^{(2)}(t,L,m_i)\approx L^{2-\eta}\,
\Upsilon_{\perp}\left({\sf t},{\sf t}_i\right)\;.
\end{equation}
In order to gain insight into the asymptotic behavior of the
scaling function $\Upsilon$,
we can use the fact that the bulk quantity $m^{(2)}(t,\infty,m_i)$
must vary $\sim t^{(2-\eta)/z}$
as $t\to\infty$ with $m_i$ fixed, where $(2-\eta)/z=1$ on the
one-loop level.
Accordingly, we should find
$\Upsilon\left({\sf t},{\sf t}_i\right)\sim {\sf t}$ in
the regime ${\sf t}_i\ll{\sf t}\ll 1$. The question then is
whether there is a distinct (anomalous)
${\sf t}$-dependence in the short-time bulk
regime ${\sf t}\ll{\sf t}_i\ll 1$. We claim that this is {\it not}
the case. The way to see this is to make a short-time expansion.
As discussed in
Refs.\ \onlinecite{jans,janszep}, the leading operator contributing
to the short-time expansion of $\phi(\bbox{x},t)$
(in correlation and response functions whose other
time arguments all stay away from zero) is
$\partial_t\phi(\bbox{x},t=0)$ because of the Dirichlet
initial condition \cite{jans,janszep}. In our case the time argument $t$
of both external points approaches zero. However, graphs contributing
to quantities such as
$\langle\partial_t\phi|_{t=0}\,\partial_t\phi|_{t=0}\rangle$
necessarily vanish since the causal structure of the theory
leaves only intervals of width zero for the integration over
internal time arguments. Hence the contribution from the one
operator in the short-time expansion of $m^{(2)}$ should give us
the limiting form of
$\Upsilon$ in the above short-time bulk regime.
This means that the
algebraic ${\sf t}$-dependence
$\Upsilon\sim{\sf t}$ should remain the same irrespective of
whether ${\sf t}_i\ll{\sf t}$ or ${\sf t}_i\gg {\sf t}$.

Detailed analytical information about the variance
$m^{(2)}_{\perp}(t)$, defined in (\ref{varitrans}), is
available in the limit $n\to\infty$.
{}From the results of Ref.\,\cite{own}
it is straightforward to show that in the bulk ($d=3$) and
for $n\to\infty$
\begin{equation}\label{largen}
m^{(2)}_{\perp} =c_1
{t}\>\frac{1+c_2\,{m}_i^2\,{t}/3}{1+c_2\,{m}_i^2\,{t}}\,,
\end{equation}
where $c_1,\,c_2$ are nonuniversal constants.
For the scaling function $\Upsilon_{\perp}({\sf t},{\sf t}_i)$, defined in
(\ref{upsitrans}), this means that
\begin{equation}\label{upslargen}
\Upsilon_{\perp}({\sf t}\ll 1,{\sf t}_i={\sf m_i}^{-z/x_i}
\ll 1) \approx {\sf t}^{(2-\eta)/z}\>
\frac{1+{\sf m}_i^2\,{\sf t}/3}{1+{\sf m}_i^2\,{\sf t}}
\end{equation}
with $(2-\eta)/z=1$ and $z/x_i=2$ in the large-$n$ limit.
Thus, in accord with the general discussion in the foregoing paragraph,
we obtain
$\Upsilon_{\perp}\approx {\sf t}$ for ${\sf t}\ll {\sf t_i}\ll 1 $
in this special case.
For ${\sf t} \approx{\sf t}_i$, on the
other hand, a crossover sets in to a limiting behavior of the form
$\Upsilon_{\perp}\approx {\sf t}/3$
for ${\sf t}_i\ll {\sf t}\ll 1$.
In the finite system this crossover should be \nopagebreak
observable as long as ${\sf t}_i<1$. For our numerical solutions,
we expect a qualitatively similar behavior,
in particular, for $n=20$.
\newpage

\def\epsfsize#1#2{0.5#1}
\hspace*{-0.2cm}\epsfbox{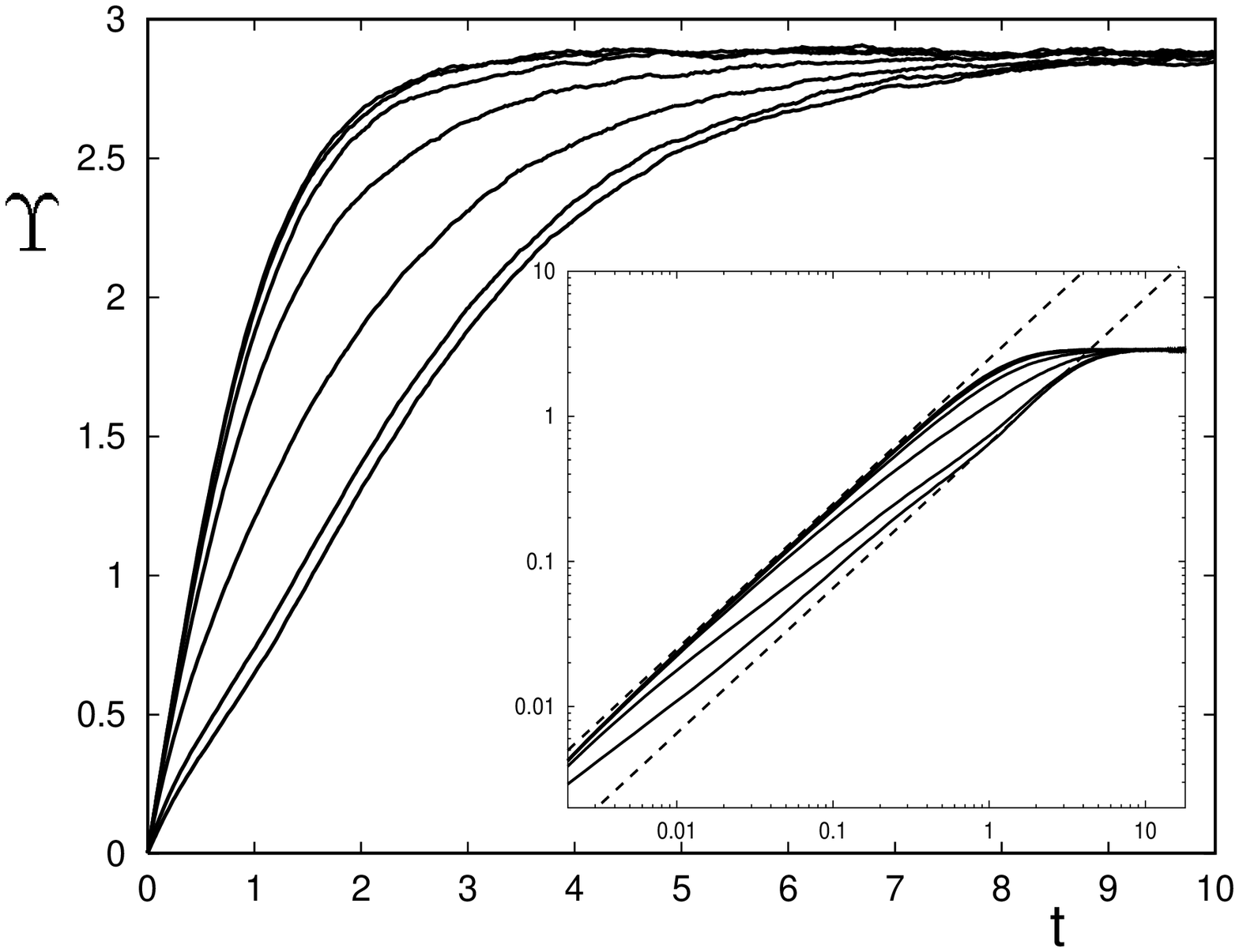}
\vspace*{-6.95cm}\\
\hspace*{10.1cm}
\noindent
\parbox{6.0cm}{{\small {\bf Fig.\,5.:}
Scaling function $\Upsilon$
for the variance defined in (\ref{upsi})
for $n=1$ and the same values of ${\sf m}_i$ used in Fig.\,2, with
${\sf m}_i$ increasing from top to bottom.
The smaller figure inserted shows the data in double-logarithmic
representation. The dashed lines indicate the bulk power laws
$\sim {\sf t}$ for
${\sf m}_i\to 0$ (upper line) and  ${\sf m}_i\to \infty$ (lower line).
}}
\\[2.6cm]

\def\epsfsize#1#2{0.495#1}
\hspace*{-0.3cm}\epsfbox{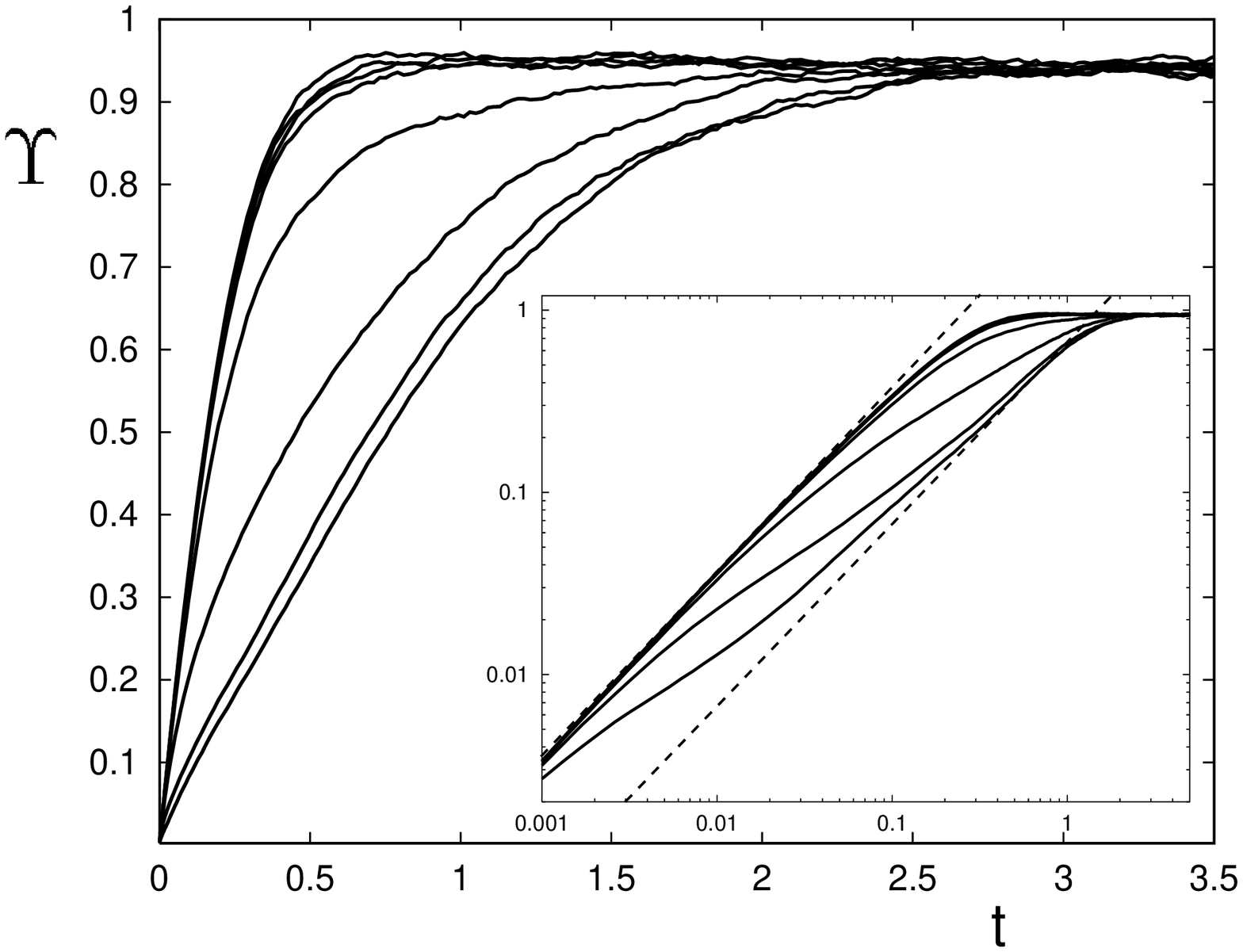}
\vspace*{-7.05cm}\\
\hspace*{10.1cm}
\noindent
\parbox{6.0cm}{{\small {\bf Fig.\,6.:}
Scaling function $\Upsilon$ defined in (\ref{upsi})
for $n=20$ and the same values of ${\sf m}_i$ used in Fig.\,3.
${\sf m}_i$ is increasing from top to bottom.
}}
\\[5.5cm]

\def\epsfsize#1#2{0.51#1}
\hspace*{-0.47cm}\epsfbox{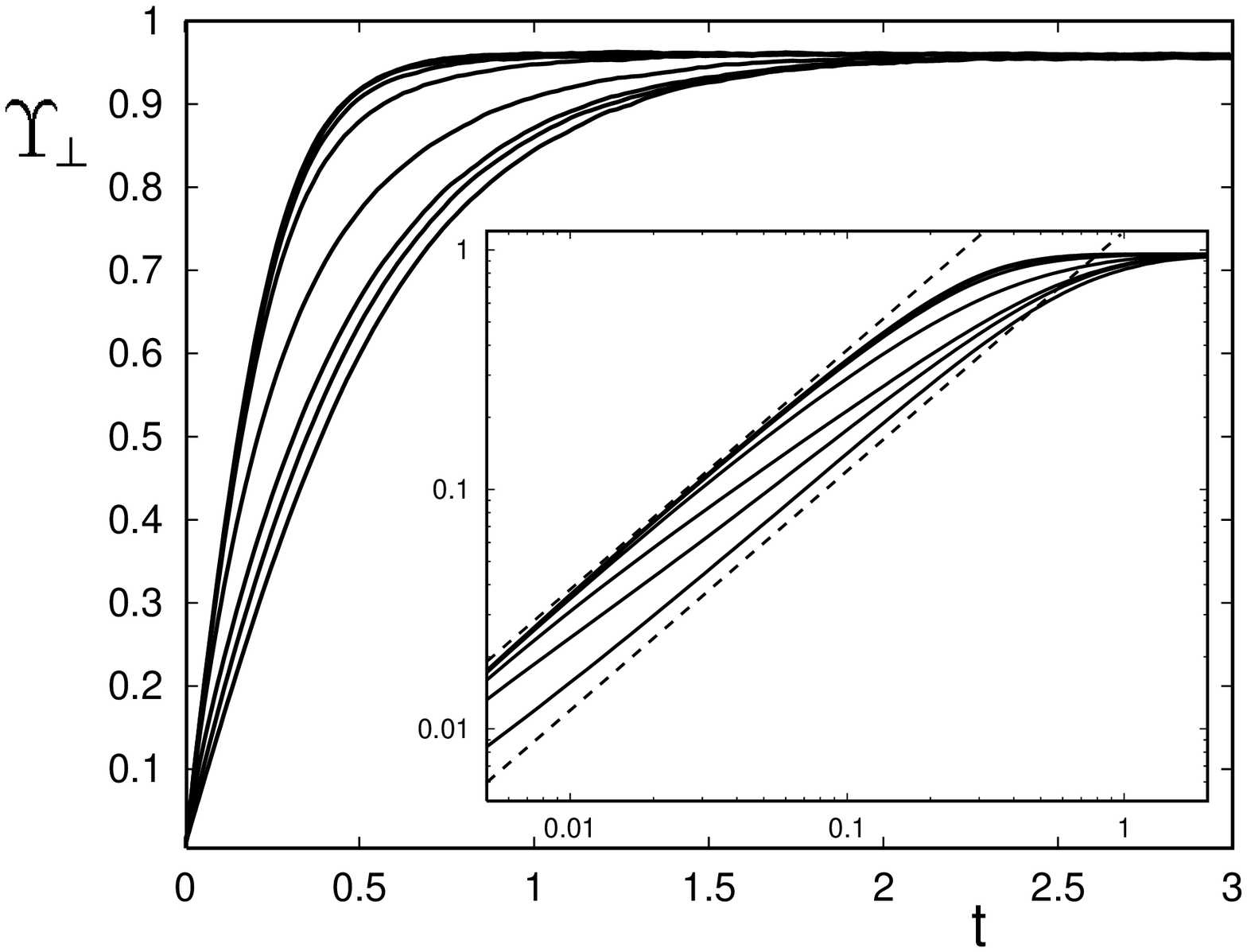}
\vspace*{-7.15cm}\\
\hspace*{10.1cm}
\noindent
\parbox{6.0cm}{{\small {\bf Fig.\,7.:}
Scaling function $\Upsilon_{\perp}$ defined
in (\ref{upsitrans})
for $n=20$. ${\sf m}_i$ is increasing from top to bottom.
Additionally to the values for ${\sf m}_i$ used in
Fig.\,3, one set of data for ${\sf m}_i=60$ has been added.
}}
\\[4.8cm]

The numerical data for $\Upsilon$ and $\Upsilon_{\perp}$
are displayed in
Fig.\,5 for $n=1$ and in Figs.\,6 and 7 for $n=20$. The results are
in conformity with the above discussion and, in the case
of $n=20$, with the exact large-$n$ result. In the log-log plots
inserted the asymptotic bulk cases $\sim {\sf t}$ are indicated
by the dashed line. The upper and lower dashed lines respresent the
limits ${\sf m}_i\to 0$
and  ${\sf m}_i\to \infty$, respectively. In Figs.\,5 and 6, these lines
have been fitted to the data. In Fig.\,7, however, the
amplitude of the lower line is one third of that of the upper line,
which demonstrates that the data for $n=20$ are
described quite well by the exact solution
(\ref{upslargen}).

\section{Concluding Remarks}

Although critical relaxation of systems with nonconserved order parameter
from a given initial state is a long-studied problem
\cite{hoha,firacz,baja,gold1,gold}, a full understanding of this
phenomenon and a theoretical description covering all
time regimes with universal behavior has only emerged recently
for bulk systems
\cite{jans,janszep,own,own2,mont,oerd,oerd2}.
In particular, serious efforts to incorporate correctly
the intial condition and
to investigate its long-time traces have only been undertaken
in the past few years.

In the present work we  have used a combination of field theory and
numerical methods to study critical relaxation in a finite system
with cube geometry and periodic boundary conditions.
Since we have taken care to build in the initial condition,
our analysis --- unlike previous approaches \cite{gold1} ---
has encompassed a proper description of the short-time
anomalous behavior of bulk systems found by Janssen et al.\ \cite{jans}
on the one hand, and on the other,
has enabled us to investigate quite generally the dependence of the
relaxational behavior on
the initial magnetization $m_i$ in the various
(macroscopic) time regimes in a systematic fashion.
Our main results are the order parameter scaling functions
$\sigma({\sf t},{\sf t}_i)$
and the scaled amplitudes $\sigma_\infty({\sf t}_i)$
depicted in Figs.\,2 and 3, and
in Figs.\,4, respectively.
Our results for $n=20$ (Fig.\,3 and 4)
are in very good agreement with the exact $n\to\infty$ results of
our previous work \cite{own}. Our results for $n=1$ display the
same qualitative behavior as those of  Ref.\ \onlinecite{mont}
obtained by Monte Carlo simulations of the three-dimensional
Glauber model. The agreement with the latter data
cannot be quantitative, of course,
simply because the one-loop values of the critical exponents
pertaining to our approximate calculation differ for $\epsilon=1$
from the $d=3$ Monte Carlo values.

Our findings corroborate that the initial condition
under favorable conditions leaves traces on long time scales.
In particular, we have been able to confirm that the amplitude of
the order parameter $m(t)$ in the regime of linear relaxational behavior
exhibits a nontrivial dependence on the initial magnetization $m_i$ of the
kind found in  Ref.\ \onlinecite{own}. Such long-time signatures of the initial
nonequilibrium state are reminiscent of aging effects
(see, e.g., Ref.\ \onlinecite{Aging} for a recent list of references). Aging
effects ---
i.e., the persistence of out-of-equilibrium effects after very long times ---
were
originally thought to occur only in systems with frustration. As pointed out
recently by Cugliandolo et al.\ \cite{Aging}, such effects appear already at
the
level of the simplest nonequilibrium systems such as the random walk or the
Langevin equation for a free scalar field. (Part of the findings of these
authors
for the free scalar field are implicitly contained already in the papers of
Janssen et
al.\ \cite{jans} and Janssen \cite{janszep}, which ---  surprisingly enough ---
were
apparently overlooked.)

Aside from being interesting from a conceptional point of view, the
improved understanding of long-time traces of the initial condition
in critical relaxation from non-equlibirum states that has been gained
may have potential benefits for numerical investigations of critical behavior.
As discussed recently by  Li et al.\ \cite{LiSchZh}, Monte Carlo simulations
of the short-time anomalous behavior seems to lend itself well to a
determination
of bulk critical exponents.

We have considered in this paper only the case of periodic boundary conditions.
It would be interesting to extend the present approach to systems with free
surfaces.
In such systems, the short-time anomalies become spatially non-homogenous. So
far there exist only two studies of short-time anomalies
in semi-infinite systems \cite{stanomalsis,RitschCzern}, which both use the
$\epsilon$ expansion
and Monte Carlo methods. None of these
studied finite-size effects.

It would also be interesting to extend our analysis to the case of quenches
beneath the critical temperature. To cope with this case,
our techniques would have to be combined with the modified handling of
the zero mode for $T<T_c$ suggested by Esser et al.\  \cite{Dohm}.\\[2mm]
\noindent
{\bf Acknowledgements:}
We greatfully acknowledge helpful discussions with
K. Oerding and H. K. Janssen.
We owe particular thanks to the former whose constructive criticism
prompted us to improve on an earlier version  of this paper.
This work was supported in part by the
Deutsche Forschungsgemeinschaft through
Sonderforschungsbereich 237.

\appendix

\section{Loop corrections to the stochastic equation %
for the zero-momentum mode}

In this appendix we consider the loop corrections to the
stochastic equation (\ref{exstocheq}) for the zero-momentum part
$\Phi(t)$. Upon insertion of the decomposition (\ref{phidecomp}), the
stochastic equation (\ref{langevin}) can be split  into its
$\bbox{q}=\bbox{0}$ and $\bbox{q}\neq\bbox{0}$ parts
\begin{equation}
\dot\phi_{\bbox{0}}^\alpha(t)\
+K_{\bbox{0}}^\alpha\left([\phi_{\bbox{0}},\psi],t\right)=\Xi^\alpha(t)
\end{equation}
and
\begin{equation}
\dot\psi^\alpha(\bbox{x},t)+K_\psi^\alpha
\left([\phi_{\bbox{0}},\psi],\bbox{x},t\right)=
\vartheta^\alpha(\bbox{x},t)\;,
\end{equation}
where $K_{\bbox{0}}^\alpha$ and $K_\psi^\alpha$ denote in abbreviated
form the deterministic ``force'' for the
$\bbox{q}=\bbox{0}$ and $\bbox{q}\neq\bbox{0}$ modes, respectively.

In order to avarage over the $\bbox{q}\neq\bbox{0}$ noise $\vartheta$,
we introduce the generating functional
\begin{equation}\label{genfun}
{\cal Z}[\Xi,J]\equiv
\left\langle\exp\left\{\int_0^{\infty}\,dt\,J(t)
\phi_{\bbox{0}}(t)\right\}\right
\rangle_{\vartheta}\,.
\end{equation}
This can be represented as a functional
integral. A standard procedure (see, e.g., Refs.\ \onlinecite{baja,zinn})
yields
\begin{eqnarray}\label{pathint}
\lefteqn{{\cal Z}[\Xi,J]\propto
\int{\cal D}\phi_{\bbox{0}}\,{\cal D}\tilde\phi_{\bbox{0}}\,
{\cal D}\psi\,{\cal D}\tilde\psi
\int{\cal D}\vartheta\,P[\vartheta]\>\times}\nonumber\\
& & \hspace*{12mm}\exp\left\{\int_0^{\infty}\,dt\,\left[
\tilde\phi_{\bbox{0}}\left(\Xi-\dot\phi_{\bbox{0}}-K_{\bbox{0}}\right)+
\tilde\psi\left(\vartheta-\dot\psi-K_\psi\right)+
J\phi_{\bbox{0}}\right]\right\}\;,
\end{eqnarray}
where $P[\vartheta]$ represents the Gaussian weight
for the $\bbox{q}\neq\bbox{0}$
noise. A spatial integration over the $\psi$ part is implied, and
the vector indices are omitted.
Contributions resulting from the functional determinant
$\det (\delta \zeta/\delta\phi)$ have
been omitted since they vanish if we use a
prepoint discretization.

The integration over
$\vartheta$ is elementary, giving
\begin{equation}\label{genfunfin}
{\cal Z}[\Xi,J]
=\int{\cal D}\phi_{\bbox{0}}\,{\cal D}\tilde\phi_{\bbox{0}}\,
{\cal D}\psi\,{\cal D}\tilde\psi
\exp\left\{-{\cal J}_{\text{eff}}+\int_0^{\infty}\,dt\,\left[J\phi_{\bbox{0}}+
\Xi\tilde\phi_{\bbox{0}}\right]\right\}
\end{equation}
with
\begin{equation}\label{effdynfu}
{\cal J}_{\text{eff}}=m_i\phi_{\bbox{0}}(0)+
\int_0^\infty dt\left\{\tilde\phi_{\bbox{0}}
\left(\dot\phi_{\bbox{0}}+K_{\bbox{0}}\right)
+\tilde\psi\left(\dot\psi+K_\psi\right)-\lambda_0\tilde\psi^2\right\}\;.
\end{equation}
{}From (\ref{genfunfin}) we see that   $\Xi$
plays the  role of a source for $\tilde\phi_{\bbox{0}}$ in
this representation. Thus ${\cal Z}[\Xi,J]$ may be
regarded as a generating functional for expectation values of
products of the operators
$\tilde\phi_{\bbox{0}}$ and $\phi_{\bbox{0}}$.
Let us introduce its Legendre transform
\begin{equation}\label{LTZ}
\Gamma [\tilde\Phi,\Phi ]=
-\ln {\cal Z}[\Xi,J]+\int_0^\infty dt\left(\Xi\tilde\Phi+J\Phi\right)\;,
\end{equation}
where it is understood that the solutions to
\begin{equation}
\tilde\Phi(t)={\delta\ln{\cal Z}[\Xi,J]\over\delta\Xi(t)}\;,
\quad \Phi(t)={\delta\ln{\cal Z}[\Xi,J]\over\delta J(t)}\;,
\end{equation}
are substituted for $\Xi$ and $J$ on the right-hand side of (\ref{LTZ}).
In terms of $\Gamma[\tilde\Phi,\Phi]$, the generating functional of
the one-particle irreducible functions pertaining to
the cumulants generated by $\ln {\cal Z}[\Xi,J]$, the desired  stochastic
equation
(\ref{exstocheq}) simply becomes
\begin{equation}\label{derieffact}
\left.\frac{\delta\Gamma[\tilde\Phi,\Phi]}{\delta \tilde\Phi(t)}
\right|_{\tilde\Phi=0}=
\Xi(t)\>.
\end{equation}

Loop corrections to the ``classical'' stochastic
equation for $\Phi(t)$
(corresponding to the substitution $\Gamma\to J_{\text{eff}}$
in (\ref{derieffact}) and given by (\ref{exstocheq}) with the
terms involving $C^{\alpha\beta}$ and $C^{\alpha\beta\gamma}$
omitted) can now be systematically computed in a standard fashion.
To this end we insert (\ref{decomp2}) and the
analogous decomposition for
$\tilde\phi_{\bbox{0}}$ into ${\cal J}_{\text{eff}}$.
{}From the
terms quadratic in $\tilde \psi,\,\psi,\delta\tilde\phi,
\,\delta\phi$
of the action (\ref{effdynfu})
we obtain the free response and correlation
propagators.
Their spatial Fourier transforms read,
in matrix notation,
\begin{equation}\label{frpp}
\bbox{G}(\bbox{q};t,t')\equiv
\left\langle\phi_{\bbox{q}}(t)\tilde\phi_{\bbox{-q}}(t')
\right\rangle_\vartheta^{\text{free}}
=\theta(t-t')\,
\exp\left\{-\lambda_0\int_{t'}^t\,dt''\,
\left[\left(\bbox{q}^2+\tau_0\right)\!\openone
+\frac{g_0}{6}\,\bbox{\cal M}(t'')\right]\right\}
\end{equation}
and
\begin{equation}\label{fcpp}
\bbox{C}(\bbox{q};t,t')\equiv
\left\langle\psi_{\bbox{q}}(t)\psi_{\bbox{-q}}(t')
\right\rangle_\vartheta^{\text{free}}
=
2\lambda_0
\int\limits_0^{\min(t,t')}dt''\,\bbox{G}(\bbox{q};t,t'')\cdot
\bbox{G}(\bbox{q};t',t'')\;,
\end{equation}
where $\bbox{\cal M}$ was defined in (\ref{matrixM}).
While (\ref{frpp}) also holds for $\bbox{q}=\bbox{0}$,
the free correlation propagator
$\langle\delta\phi(t)\delta\phi(t')
\rangle_\vartheta^{\text{free}}$ vanishes
identically. The latter is due to the absence of
a term
quadratic in $\tilde\phi_{\bbox{0}}$ in
${\cal J}_{\text{eff}}$, which in turn
is a direct consequence of the
fact that we averaged only over the $\bbox{q}\ne\bbox{0}$
part $\vartheta$ or the noise, but not over its
$\bbox{q}=\bbox{0}$ part $\Xi$.

Up to two-loop order the functions ${\cal C}^{\alpha\beta}(t)$
and ${\cal C}^{\alpha\beta\gamma}(t)$
in (\ref{exstocheq}) are given by
\def\epsfsize#1#2{0.6#1}
\begin{eqnarray}
{\cal C}^{\alpha\beta}(t)& = & \epsfbox{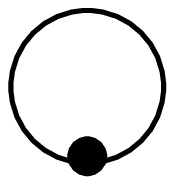}+
\epsfbox{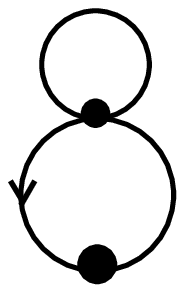}+{\cal O}(\text{3-loops})
\end{eqnarray}
and
\begin{equation}
{\cal C}^{\alpha\beta\gamma}(t)=\epsfbox{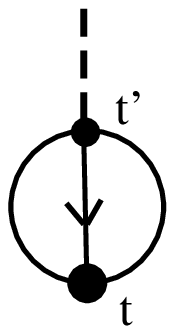}+
{\cal O}(\text{3-loops})\,,
\end{equation}
where the full lines with and without an arrow denote
the free response and correlation propagators (\ref{frpp}) and
(\ref{fcpp}), respectively, while the broken line signifies $\Phi(t)$.

The explicit expressions for the graphs are
\begin{equation}\label{ce1}
\epsfbox{stodia1.eps}=C^{\alpha\beta}(t,t)=
L^{-d}\sum_{\bbox{q}\ne\bbox{0}}
C^{\alpha\beta}(\bbox{q};t,t)\;,
\end{equation}
\begin{equation}\label{ce2}
\epsfbox{stodia2.eps}=\lambda_0 g_0 L^{-d}\int\limits_0^\infty dt'
\sum_{\bbox{q}\ne\bbox{0}}
C^{\alpha\alpha'}(\bbox{q};t,t')\,G^{\beta\beta'}(\bbox{q};t,t')\,
S^{\alpha'\beta'\gamma\delta}\,{\cal C}^{\gamma\delta}(t')
\end{equation}
and
\begin{equation}\label{ce3}
\epsfbox{stodia3.eps}=3\lambda_0 g_0 L^{-2d}\,
S^{\alpha'\beta'\gamma'\delta'}
\int\limits_0^\infty dt'\,\Phi^{\delta'}(t')
\sum_{\bbox{q},\bbox{q}'\ne\bbox{0}\atop
\bbox{q}+\bbox{q}'+\bbox{q}''=\bbox{0}}
C^{\alpha\alpha'}(\bbox{q};t,t')\,
C^{\beta\beta'}(\bbox{q}';t,t')\,
G^{\gamma\gamma'}(\bbox{q}'';t,t')\;.
\end{equation}

{}From expression (\ref{ce1})
the one-loop result (\ref{ceij}) for $\bbox{\cal C}$
used in the main text follows in a straightforward manner.

\newpage
\section{Laurent expansion of required integrals}

Equations (\ref{tau0t}) and (\ref{g0t}) involve the integrals
\begin{equation}
I_p(t)\equiv \int_0^tds\,s^p\,
\left[A^d(\kappa s)-1\right]e^{-\rho \kappa s}
\end{equation}
with
$\kappa =8\pi^2\lambda_0L^{-2}$, $\rho=L^2\tau_0/4\pi^2$, and
$p=0,1$. Here we will compute the Laurent expansion of these
to ${\cal O}(\epsilon^0)$.

{}From $A^d(\kappa s)$ we add and
subtract its limiting form $(\kappa s/\pi)^{-2+\epsilon}$
for $s\to 0$. This gives
\begin{eqnarray}
I_p(t)&=&\label{Ip}
\kappa^{-(p+1)}\left[\pi^{d/2} J_p(\kappa t,\rho)+
F_p(\kappa t,\rho)+{\cal O}(\epsilon)\right],
\end{eqnarray}
where $F_p$ is the function defined in (\ref{Fp}), while
\begin{equation}
J_p(\upsilon,\rho)\equiv \int_0^\upsilon \!dx\,
x^{p-2+\epsilon/2}\,e^{-\rho x}=
\Gamma(p-1+\epsilon/2)\,\upsilon^{p-1+\epsilon/2}\,
\gamma^*(p-1+\epsilon/2,\upsilon \rho)\;.
\end{equation}
Here $\gamma^*(a,x)=x^{-a}\gamma(a,x)/\Gamma(a)$ is the
analytic function related to the incomplete gamma
function $\gamma(a,x)$ \cite{GradRyz}. The poles arise solely from $J_p$.
Therefore we have expanded the second (regular) term inside the brackets.

The Laurent expansion of the function $J_p$ can be obtained in a
straightforward fashion, utilizing
the recursion relation
\begin{equation}
\gamma^*\!\left(-1+{\epsilon\over 2},x\right)=
x\,\gamma^*\!\left({\epsilon\over 2},x\right)
+{\exp(-x)\over\Gamma(\epsilon/2)}\;,
\end{equation}
familiar properties of $\Gamma(a)$, and
the expansion \cite{Abramowitz}
\begin{equation}
\gamma^*\!\left({\epsilon\over 2},x\right)=1+
{\epsilon\over 2}\,\Big[\text{Ei}(-x)-\ln x\Big]+
{\cal O}(\epsilon^2)\;,
\end{equation}
where the exponential-integral
function $\text{Ei}(x)$ of Ref.\ \onlinecite{GradRyz} used here is
related to the function E$_1$ of Ref.\ \onlinecite{Abramowitz}
via $\text{Ei}(-x)=-\text{E}_1(x)$.

One finds
\begin{equation}
J_0(\upsilon,\rho)=\rho\left[\frac{-2}{\epsilon}+C_E-1+\ln \rho
-\text{Ei}(-\upsilon\rho)-\frac{\exp(-\upsilon\rho)}{\upsilon\rho}\right]
+{\cal O}(\epsilon)
\end{equation}
and
\begin{equation}
J_1(\upsilon,\rho)=\frac{2}{\epsilon}+
\text{Ei}(-\upsilon\rho)-\ln\rho-C_E
+{\cal O}(\epsilon)\;.
\end{equation}
Substitution of these results into (\ref{Ip}) leads to
\begin{eqnarray}\label{I0}
I_0(t)&=&\lambda_0^{-1}K_d\,L^2\left\{L^2\tau_0
\left[\frac{-2}{\epsilon}-
\ln\frac{2\lambda_0 t}{L^2}-B(2\lambda_0\tau_0t)-
\frac{\exp(-2\lambda_0\tau_0 t)}{2\lambda_0\tau_0 t}\right]\right.\nonumber\\
&&\left.\mbox{}\qquad\qquad+
F_0\!\left(\frac{8\pi^2\lambda_0t}{L^2},
\frac{L^2\tau_0}{4\pi^2}\right)+{\cal O}(\epsilon)\right\}
\end{eqnarray}
and
\begin{eqnarray}\label{I1}
I_1(t)&=&\left(\frac{L^2}{8\pi^2\lambda_0}\right)^2\,\left\{
\pi^{d/2}\left[\frac{2}{\epsilon}+B(2\lambda_0\tau_0t)
+\ln L^2\tau_0-C_E\right]\right.
\nonumber\\&&\left.\qquad\mbox{}
+F_1\!\left(\frac{8\pi^2\lambda_0t}{L^2},
\frac{L^2\tau_0}{4\pi^2}\right)
+{\cal O}(\epsilon)
\right\}
\end{eqnarray}

\end{document}